\documentclass[usenatbib]{mn2e} 
\usepackage{psfig}
\usepackage{lscape}
\usepackage{rotating}
\usepackage{psfig}
\usepackage{amsmath}
\usepackage{amssymb}
\title[Emission line gas ionisation in young radio galaxies]{The 
ionisation of the emission line gas in young radio galaxies}
\author[J. Holt et al.]{J. Holt$^{1}$\thanks{E-mail:
jholt@strw.leidenuniv.nl}, C. N. Tadhunter$^{2}$ and R. Morganti$^{3,4}$\\
$^{1}$Leiden Observatory, Leiden University, PO Box 9513, 2300 RA
  Leiden, The Netherlands.\\
$^{2}$Department of Physics and Astronomy, University of Sheffield,
  Sheffield, S3 7RH, UK.\\
$^{3}$Netherlands Institue for Radio Astronomy, Postbus 2,
7990 AA Dwingeloo, The Netherlands.\\
$^{4}$Kapteyn Astronomical Institute, University of Groningen, Postbus
800, 9700 AV Groningen, The Netherlands.}

\usepackage{ulem}

\newcommand{\kms}{km s$^{-1}$}
\newcommand{\ha}{H$\alpha$}
\newcommand{\hb}{H$\beta$}

\newcommand{\ebv}{E(B-V)}
\newcommand{\lala}{$\lambda\lambda$}
\newcommand{\dipso}{erg s$^{-1}$ cm$^{-2}$ \AA$^{-1}$}
\defcitealias{villar99a}{Villar-Mart{\'{i}}n et al., 1999a}

\begin{document}
\maketitle
\begin{abstract}
This paper is the second in a series in which we present
intermediate-resolution, wide-wavelength coverage spectra for a
complete sample of 14 compact radio sources, taken with the aim of
investigating the impact of the nuclear activity on the cirumnuclear
interstellar medium (ISM) in the early stages of radio source
evolution. In the first paper \citep{holt08} we presented the
kinematic results from 
the nuclear emission line modelling and reported fast outflows in the
circumnuclear gas. In this paper, we use the line fluxes to
investigate the physical conditions and dominant ionisation mechanisms
of the emission line gas. We find evidence for large electron
densities and high reddening in the nuclear regions, particularly in
the broader, blueshifted components. These results  are consistent with the idea
that the young, recently triggered radio sources still reside in 
dense and dusty cocoons deposited by the recent activity triggering
event (merger/interaction). In addition, we find that the quiescent
nuclear and extended narrow components 
 are consistent with AGN photoionisation, split between simple-slab AGN
photoionisation and mixed-medium photoionisation models. For the
nuclear broader  and shifted components the results are less
clear. Whilst there are suggestions that the broader components may be
closer to shock plus precursor models on the diagnostic diagrams, and
that the electron temperatures and densities are high, we are unable
to unambiguously distinguish the dominant ionisation mechanism using
the optical emission line 
ratios. This is surprising given the strong evidence for jet-cloud
interactions  (broad emission lines, large
outflow velocities and strong radio-optical alignments), which favours
the idea that the warm gas has been accelerated in shocks driven by
the radio lobes expanding through a dense cocoon of gas deposited
during the triggering event.
\end{abstract}


\section{Introduction}
It is now understood that AGN-feedback, in the form of outflows in the
circumnuclear ISM, is likely to be a crucial part of galaxy evolution
and is responsible for the close relationship between the black hole
mass and the galaxy bulge properties
(e.g. \citealt{ferrarese00,gebhardt00,tremaine02,marconi03}). However,
due to a lack of observational evidence, feedback is often invoked as
a `black box' and 
often relies on powerful AGN winds to drive the outflows and shed
the natal cocoon deposited during the triggering event 
(e.g. merger/interaction; \citealt{dimatteo05,hopkins05}). Whilst
this may be true for radio-quiet AGN, it is likely that in radio-loud
AGN, the expanding radio jets also 
contribute to the feedback through jet-induced outflows
(e.g. \citealt{bicknell97}), particularly during the early stages of
the radio source evolution when the radio source and ISM are  on
similar scales. 

Recent work (by e.g. \citealt{best06,bower06} \&
  \citealt{croton06}) has shown that extended radio sources have a
  significant impact on the hot gas on cluster scales through feedback effects
  (e.g. the quenching of cooling flows in massive haloes). Hence, 
a significant contribution to the overall AGN
feedback from the young, expanding radio sources on smaller scales should be
expected.   
Indeed, over the last few years, fast nuclear outflows (up to $\sim$
2000 \kms) have been 
detected in a number of compact (young) radio sources, both in the
optical emission lines
(e.g. \citealt{tadhunter01,holt03,holt06,holt08}; the latter 3 references: H03,
  H06 \& H08 hereafter) and in neutral
hydrogen as deep absorption features
(e.g. \citealt{morganti03a,morganti04,morganti05}).  

In H08, we reported  results
for the emission line kinematics in a 
complete sample of 14 compact radio sources, including both Compact
Steep Spectrum radio sources (CSS: size $<$ 15kpc) and Gigahertz-Peaked
Spectrum radio sources (GPS: size $<$ 1 kpc).
 The observed line widths and blueshifts (up to $\sim$
2000 \kms) in the nuclear emission line gas are too large to be
explained by gravitational motions. Instead, the extreme kinematics were
interpreted as signatures of strong interactions between the young,
small-scale radio jets and the emission line gas, following the
 model proposed for PKS 1549-79 by \citet{tadhunter01}. This
interpretaion is supported by evidence  for  strong
radio-optical alignments in {\it all} CSS sources with deep HST
imaging, including some 
sources in this sample
(e.g. \citealt{devries97,devries99,axon00,privon08}). 

Although the expanding radio jets provide a convenient outflow driving
mechanism, consistent with the  results on the emission line
kinematics, it is important to 
  test this scenario further. In this paper we use  optical emission line
ratios (from the line modelling in H08) 
to investigate the physical 
conditions and the dominant ionisation mechanism(s) of the narrow line
gas. In particular  to determine whether the emission line regions are
shocked, as expected in the case that  the outflows are being accelerated
by jet-cloud interactions
(e.g. \citealt{villar98,clark98,solorzano01}).  In Section 2 we
give a brief overview of the sample, observations, data reduction and
analysis techniques and refer the reader to H08 for a detailed
discussion. In Section 3 we discuss the line fluxes and, in particular, the
reddening, density and temperature of the emission line gas,  giving  brief
details of the line modelling.
We discuss the line ratios in detail in Section 4, along with
diagnostic diagrams and radio-optical correlation plots. All line
fluxes, aperture information and extracted nuclear spectra are presented
in the Appendices. Information on the emission line kinematics for all
sources is presented in H08. 

Throughout this paper we assume the following cosmology: H$_{0}$ = 71
\kms, $\Omega_{\rmn 0}$ = 0.27 and  $\Omega_{\Lambda}$ = 0.73.

\section{Sample Selection, Observations, Data reduction and Analysis
  Techniques} 
This paper is the second in a series of two reporting  deep optical
spectroscopic observations of a complete sample of compact radio
sources. Below we summarise the important points and refer readers to 
H08 for a detailed description 
of the sample, observations, data reduction and analysis techniques.

Our complete  sample comprises 14 compact radio sources derived from
the northern 3C and 4C samples and southern 2Jy sample and includes 8 CSS, 3 GPS,
2 compact core sources and 1 compact flat-spectrum source. The sample
has intermediate redshifts ($z$ $<$ 0.7) and a radio power range of 26 $<$ log P$_{\rmn 5GHz}$
$<$ 28  (W Hz$^{-1}$). 

Long-slit optical spectroscopic observations of the
full sample were obtained during three observing runs with ISIS/WHT,
EMMI/NTT and FORS2/VLT. In order to include the outflowing regions in the
slit, and to ensure the spectra were of sufficient resolution, the 
 observations were made using a 
 1.0-1.5 arcsec slit, while to reduce the
 effects of differential atmospheric refraction, all objects were
 observed either at low airmass (sec $z$ $<$ 1.1) and/or with the slit
 aligned along the parallactic angle. Due to various observational
 constraints, we have only aligned the slit along the radio axis for
 approximately half of the sources.  For full details of the observational
 setup, we refer readers to Table 2 of H08.

The data were reduced in the usual way (bias subtraction, flat
fielding, cosmic ray removal, wavelength calibration, flux
calibration) using the standard packages in {\sc iraf}. To ensure good
flux calibration ($\pm$5 per cent), several spectrophotometric
standard stars were observed during each run with a wide (5 arcsec)
slit. This accuracy was confirmed by good matching in the flux between
the red and blue spectra. \footnote{The only exception was 3C
  303.1. For this source the only ratio data used is for lines observed
  with the same arm of the ISIS spectrograph.} Additional standard
stars were 
observed with a narrow slit to correct the spectra for atmospheric
absorption features (e.g. A and B bands). The spectra of all 
sources were corrected for Galactic extinction using the \ebv~values
from \citet{schlegel98} and the \citet{seaton79} extinction law.
The wavelength calibration accuracies were 0.06-0.53\AA, 
0.06-0.24\AA~and 0.20-0.24\AA~for the WHT, NTT and VLT data
respectively, dependent on  
wavelength range. Similarly, the spectral resolutions were
typically 3.3-4.8$\pm$0.2\AA, 4.3-6.7$\pm$0.1\AA~and 6.5-7.4\AA~
for the WHT, NTT and 
VLT data, again dependent on wavelength range. The spatial resolutions
were 0.36 arcsec/pixel, 0.33 arcsec/pixel and 0.25 arcsec/pixel for the
WHT, NTT and VLT data respectively. The seeing range was 0.8-2.5
arcsec. Full details are given in  Table 2 of H08.

As discussed in H08, in order to detect and accurately model the
broader emission line components, it is necessary to model and
subtract the underlying continuum emission. For the broader, blueshifted emission
line components, variations in the continuum can make it difficult to
establish the centroid, FWHM and line flux of the Gaussian. If the
underlying continuum also contains strong stellar absorption lines
(e.g. \hb~etc), a 
good model of the continuum is required to ensure that the emission line
fluxes are accurately modelled. \citet{holt05} estimated that
subtracting different continuum models which all provided a reasonable
fit to the SED can cause a difference of up to 10\% in the flux in
the broader components, although typically the narrower
components are less affected.

As a first step in modelling the continuum, we generated and
subtracted  the nebular continuum following \citet{dickson95}, taking
full account of reddening following the techniques outlined in H03. 
The remaining
continuum was then modelled  using a customised {\sc idl}
minimum $\chi^2$ fitting programme (see
e.g. \citealt{robinson01,tadhunter05,holt07} for details) allowing up
to three separate continuum components: 12.5 Gyr unreddened Old
Stellar Population (OSP), a Young Stellar Population (YSP)
with reddenings 0 $<$ \ebv~$<$ 1.6 and age 0.01 $<$ $t_{\rmn YSP}$ $<$ 5.0 Gyr,
and an AGN/power law component. The best fitting model, defined as that with
the least number of components required to adequately model both the
overall SED and discrete stellar absorption features
(e.g. Ca H+K, Balmer series),
 was then subtracted. The continua of three sources were not
modelled in this way: 3C 277.1 (quasar), PKS 1814-63 (foreground star) and 3C
303.1 (mis-matching of the blue and red spectra).  Full details of the subtracted
models can be found in H08. 

The spectra were extracted and analysed using the {\sc starlink}
packages {\sc figaro} and {\sc dipso}. Note, some components for a few
key diagnostic lines were not detected in some apertures. For these, we have
estimated upper limits for the flux by adding Gaussians to the data
until the line component could just be detected on visual inspection
of the spectra.

Throughout this paper we use the kinematic component definition
defined in H08: 
\begin{itemize}
\item narrow: FWHM $<$ 600 \kms;
\item intermediate: 600 $<$ FWHM $<$ 1400 \kms;
\item broad: 1400 $<$ FWHM $<$ 2000 \kms;
\item very broad: FWHM $>$ 2000 \kms.
\end{itemize}

\section{Results}
As discussed above, the nuclear emission lines in
compact radio sources are often broad, with asymmetric profiles
requiring multiple Gaussian components to model them. In H08, we
modelled the nuclear emission lines for all sources in our sample,
identifying between 2 and 4 Gaussian components with velocity widths
and blueshifts of up to 2000 \kms. Here, we use the emission
line models from H08 and focus our analysis on the emission line
fluxes and ratios. We use emission line ratios and diagnostic diagrams to
investigate whether the compact radio sources in our sample still
retain their dense and dusty natal cocoons (reddening and emission
line density), and to investiage the dominant ionisation mechanism(s) and
the implications for the likely outflow driving mechanism. For
brevity, the line fluxes of all lines in all apertures extracted are
presented in Appendix A. 

\subsection{Aperture extraction and emission line modelling}
For a detailed discussion of the emission line modelling, we refer
readers to H08 and we only summarise the key points here. 

\subsubsection{Aperture selection}
In H08, with the exception of modelling the most extended emission
lines in the spatial direction to accurately  determine the systemic
velocities, the main  results on the emission line kinematics were derived from wide
nuclear apertures centred on the nuclear continuum emission. These
nuclear apertures, which also form the basis of the analysis in this
paper,  are shown in Appendix \ref{appendix2} along with
spatial profiles of the bright emission lines and the continuum
emission, and the full extracted nuclear spectra. In addition, 5
sources (3C 268.3, 3C 277.1, PKS 1345+12, PKS 1814-63 \& 3C 459) show
evidence for extended emission 
in several lines and we have extracted further apertures. The
positions of these, along with the extracted spectra, are also shown
in  Appendix B.

\subsubsection{Emission line modelling}
In general, we have modelled all emission lines in a particular
aperture using a kinematic model derived from the {[O 
    III]}\lala4959,5007 doublet, a standard technique for modelling the
emission lines in radio galaxies. In short, the
velocity widths and shifts of the various components were fixed to be
the same as for the {[O III]} doublet, with further constraints
arising from atomic physics.  As all lines can be reproduced by
the same model, we can assume with confidence that the emission
originates from similar regions of the ISM. Hence, it is meaningful to
use emission line ratios to investigate the physical and ionisation
properties of the emission line gas. 

Notable exceptions are the nuclear apertures of the GPS sources PKS 
1345+12 and PKS 1934-63. For these apertures, one model did not
reproduce all lines. In these latter cases, only components
consistent in all lines (e.g. the narrow components in PKS 1345+12)
are plotted in the diagnostic diagrams.  Finally, for 3C 
303.1, we only consider line ratios from lines in the same arm of the
instrument as the two arms do not match in flux.

\subsection{Emission line densities}
\label{sect:density}
During the early stages of radio source evolution, the nuclear regions
of radio sources are likely to harbour a rich ISM, deposited during
the activity triggering event.
Here, we investigate the density of the
circumnuclear gas using the  {[S
    II]} 6716/6731 density diagnostic ratio.
The measured ratios were converted to densities
using the {\sc iraf} program {\sc temden}, part of the {\sc stsdas}
package, which is based on the five-level atom calculator developed by
\citet{derobertis87}. For all calculations, we have assumed an
electron temperature of T$_{\rmn e}$ = 10,000K, although the {[S II]}
ratio varies little with temperature.
As discussed above,  when modelling the emission lines in
the nuclear aperture of each source, the profiles of all lines were
well fitted by the corresponding {[O III]} model. The only exceptions
were PKS 1345+12 (discussed in detail in H03) and PKS
1934-63. 

\begin{figure}
\centerline{\psfig{file=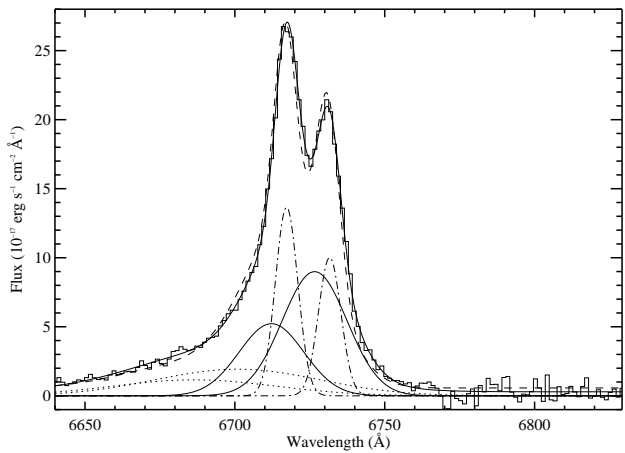,width=9cm,angle=0.}}
\caption[Example of SII doublet]{The {[S II]}\lala6716,6731 doublet in
  the nuclear aperture of PKS 1345+12 (taken from H03). The faint
  solid line (histogram) represents the 
  data, the bold line is the overall best fitting model to the
  doublet, the dashed line is the {[O III]}
  model to the {[S II]} 
  doublet and the dot-dashed, solid and dotted pairs of lines are the
  narrow, intermediate and broad components. This plot highlights the
  difficulties in obtaining a unique model for the {[S II]}
  doublet. Although it is relatively straightforward to fit the correct
  number of components, 
  and we can confidently determine the {\it total flux} in, for example, the
  intermediate or broad components,  
  the widths and shifts of the broader components are comparable to
  the separation of the lines in the doublet. This  makes it difficult to
  determine the ratio between the two intermediate or two broad
  components, which is critical for determining the density.}
\label{fig:s2}
\end{figure}

The derived electron densities are summarised in Table
{\ref{tab:physcond}}. In the nuclear apertures, we have been able to
accurately determine the electron density for 5 components only: 3C
213.1 (narrow), 3C 303.1 (intermediate), PKS 0023-26 (narrow), PKS
1934-63 (narrow) and 3C 459 (narrow). For the remaining components, we
have only been able to determine whether the components are consistent
with low or high density. The reason for this is that 
the blend is highly complex,
requiring 4 to 8 Gaussian components (2-4 per line) for an adequate
fit, and the velocity 
widths and shifts of the broader components are often comparable to the
separation of the lines in the doublet (see Figure {\ref{fig:s2}} and
also H03); it is usually possible to confidently determine the {\it
  total flux} in e.g. the broad components together, but to determine
the density, the ratio is required. 
Hence, whilst the {[O III]} model generally provides a good
fit to the {[S II]} doublet, in many cases it is not possible 
to obtain a {\it physically viable} good fit without forcing the
6716/6731 ratio to the high (0.44) or low (1.42) density limit of the
diagnostic.

Although we have not been able to accurately measure the electron
densities in the majority of apertures, from the results presented in
Table {\ref{tab:physcond}}, we can draw the following general
conclusions:
\begin{enumerate}
\item The nuclear narrow components tend to be consistent with low
  electron densities (few hundred cm$^{-3}$), as do the extended
  narrow components.
\item The nuclear broader components (intermediate/broad/very broad)
  appear to be consistent with higher densities (n$_{e}$ $>$ 10$^3$
  cm$^{-3}$).  
\end{enumerate}

\begin{table*}
 \begin{minipage}{26cm}
  \vspace*{4cm}
   \rotcaption[Physical conditions.]{\footnotesize Summary of the
     physical conditions (density, reddening \& temperature)
   for the sample. Where there are two   
components with similar FWHM (i.e. two narrow components), the
component at the systemic velocity is 
presented first (for PKS 1306-09 the blue narrow component is presented first).
$\dagger$ was measured and found to be consistent with either the low density limit
({[S II]}$\lambda$6716/$\lambda$6731 = 1.42) or the high density limit
({[S II]}$\lambda$6716/$\lambda$6731  
= 0.44), within the uncertainties.  
$\ddagger$ forced the the low (1.42) or high (0.44) density limit. See
text for details. $\star$ sources 
in which there is a clear trend of increasing density (column $(g)$) or
reddening (column $(l)$) 
  with FWHM. Blank positions in the table
indicate kinematic components not observed in a source/aperture. For
components which are not detected in a relevant diagnostic line, the
entry is {--}. }
\label{tab:physcond}
\hspace{9cm}
\vspace{9cm}
  \begin{rotate}{90}
   \centering
    \hspace{0cm}%
{\footnotesize
\begin{tabular}{l l l l l l l l l l l l} \hline\hline 
 & & \\
\multicolumn{1}{l}{Object}  &\multicolumn{1}{l}{Aperture} & 
\multicolumn{5}{l}{\bf Electron density, n$_e$ (cm$^{-3}$)} &
\multicolumn{5}{l}{\bf E(B-V) value}\\
 & &\multicolumn{1}{l}{n} & \multicolumn{1}{l}{i} &
\multicolumn{1}{l}{b} & \multicolumn{1}{l}{vb} & & n & i & b & vb & \\
\multicolumn{1}{l}{$(a)$}&\multicolumn{1}{l}{$(b)$}&
\multicolumn{1}{l}{$(c)$}&\multicolumn{1}{l}{$(d)$}&
\multicolumn{1}{l}{$(e)$}&\multicolumn{1}{l}{$(f)$}&
\multicolumn{1}{l}{$(g)$}&\multicolumn{1}{l}{$(h)$}&
\multicolumn{1}{l}{$(i)$}&\multicolumn{1}{l}{$(j)$}&
\multicolumn{1}{l}{$(k)$}&\multicolumn{1}{l}{$(l)$}\\
\hline\hline
3C 213.1    & nuc  &\multicolumn{1}{l}{480$_{-150}^{+180}$} &
\multicolumn{1}{l}{--} &&&& zero & $>$0.2 & & &$\star$ \\
3C 268.3    & nuc  & 1.42$\ddagger$/1.42$\dagger$                   &
\multicolumn{1}{l}{1.42$\ddagger$} &&&& zero/zero & 0.7$^{+0.1}_{-0.1}$
& & &$\star$\\
            & SE       &\multicolumn{1}{l}{1.42$\dagger$}&&&&&zero \\
            & NW       &\multicolumn{1}{l}{--}&&&&&0.6$^{+0.3}_{-0.2}$ \\
3C 277.1    & nuc  & \multicolumn{1}{l}{$<$600} &
\multicolumn{1}{l}{0.44$\ddagger$} & & && zero & 0.7$^{+0.3}_{-0.4}$ &
& BLR:zero & $\star$\\
            & SE blob  & --/-- & & & && zero/zero \\
            & SE HII   & \multicolumn{1}{l}{--} & & && & zero\\
4C 32.44    & nuc  & --/-- & & \multicolumn{1}{l}{--} &
\multicolumn{1}{c}{--} & & zero/zero & & 0.9$^{+0.1}_{-0.1}$&1.0$^{+0.3}_{-0.6}$&$\star$ \\
PKS 1345+12 & nuc  & \multicolumn{1}{l}{1.42$\ddagger$ ($<$150)} &
\multicolumn{1}{l}{$>$ 5300} & \multicolumn{1}{l}{$>$ 4200} &
&{$\star$} & zero & 0.4$^{+0.1}_{-0.1}$ & 1.4$^{+0.5}_{-0.5}$ & &$\star$\\
            & NW,PA160 & \multicolumn{1}{l}{500$_{-200}^{+300}$} & & &
& &zero &\\
            & SE,PA160 & 1.42$\dagger$/1.42$\ddagger$ & && & &
0.5$^{+0.1}_{-0.1}$/1.2$^{+0.4}_{-0.7}$ &\\
            & NE,PA230 & \multicolumn{1}{l}{1.42$\dagger$} & && & &
0.7$^{+0.2}_{-0.3}$ &\\
            & SW,PA230 & --/1.42$\dagger$ & & & & & --/0.7$^{+0.1}_{-0.1}$&\\
3C 303.1    & nuc & \multicolumn{1}{l}{1.42$\ddagger$} &
\multicolumn{1}{l}{110$_{-40}^{+40}$} & & &{$\star$}? & zero &
0.5$^{+0.2}_{-0.2}$ && &$\star$\\
PKS 0023-26 & nuc  & \multicolumn{1}{l}{400$_{-150}^{+200}$} &
\multicolumn{1}{l}{1.42$\dagger$} & & & & zero & zero &  \\
PKS 0252-71 & nuc  &  & & & & & zero/zero & zero &\\
PKS 1306-09 & nuc  &  & & & & & zero/zero &\\
PKS 1549-79 & nuc  & \multicolumn{1}{l}{1.42$\dagger$} &
\multicolumn{1}{l}{0.44$\ddagger$} & & & & zero & zero & &
BLR:2.8$^{+0.1}_{-0.1}$ &  \\
            & PA75    & \multicolumn{1}{l}{1.42$\dagger$}
&\multicolumn{1}{l}{0.44$\ddagger$}  & & & & 0.4$^{+0.1}_{-0.1}$ &
0.4$^{+0.1}_{-0.2}$ &\\
            & PA-5    & \multicolumn{1}{l}{1.42$\dagger$} & & & & & --
&\\
PKS 1814-63 & nuc  & 1.42$\dagger$/700$_{-150}^{+150}$ & & & & & --/--\\
            & ext & \multicolumn{1}{l}{240$_{-90}^{+100}$}\\
PKS 1934-63 & nuc  & \multicolumn{1}{l}{600$_{-200}^{+300}$}& &
\multicolumn{1}{l}{0.44$\ddagger$} & & & zero \\
PKS 2135-20 & nuc  & &   --/-- & & & & & zero/1.3$^{+0.4}_{-0.2}$ & \\
3C 459      & nuc  & \multicolumn{1}{l}{300$_{-100}^{+100}$} & &
\multicolumn{1}{l}{0.44$\ddagger$} & & & zero & & 1.3$^{+0.1}_{-0.1}$
&& $\star$ \\
            & ext & \multicolumn{1}{l}{1.42$\ddagger$} & &
\multicolumn{1}{l}{1.42$\ddagger$}&&&zero && zero  \\
\hline\hline
\end{tabular}
}
  \end{rotate}
 \end{minipage}
\end{table*}
\setcounter{table}{0}
\begin{table*}
\caption[Physical conditions.]{{\it continued.}}
\label{tab:physcond}
{\footnotesize
\begin{tabular}{l l l l l l l} \hline\hline 
 \\
\multicolumn{1}{l}{Object}  &\multicolumn{1}{l}{Aperture} & 
\multicolumn{4}{l}{\bf Electron temperature, 10$^{3}$ T$_e$ (K)} \\
 & &\multicolumn{1}{l}{n} & \multicolumn{1}{l}{i} &
\multicolumn{1}{l}{b} & \multicolumn{1}{l}{vb}\\
\multicolumn{1}{l}{$(a)$}&\multicolumn{1}{l}{$(b)$}&
\multicolumn{1}{l}{$(l)$}&\multicolumn{1}{l}{$(m)$}&
\multicolumn{1}{l}{$(n)$}&\multicolumn{1}{l}{$(o)$}\\ 
\hline\hline
3C 213.1    & nuc  &\multicolumn{1}{l}{$<$19} &
\multicolumn{1}{l}{$<$30}\\
3C 268.3    & nuc  &
16$^{+8}_{-3}$/15$^{+4}_{-2}$ &
18$^{+18}_{-4}$ \\ 
            & SE       &18$^{+1}_{-1}$\\
            & NW       &\multicolumn{1}{l}{--}\\
3C 277.1    & nuc  & 25$^{+1}_{-1}$ &
50$_{-20}^{+100}$&&&\\
            & SE blob  & $<$35/$<$55\\
            & SE HII   & \multicolumn{1}{l}{$<$95}\\
4C 32.44    & nuc  & 11$^{+6}_{-2}$/$<$18 & & \multicolumn{1}{l}{$<$13} &
\multicolumn{1}{c}{$<$70}\\
PKS 1345+12 & nuc  &\\
            & NW,PA160 & \multicolumn{1}{l}{14$_{-2}^{+3}$}\\
            & SE,PA160 &\\
            & NE,PA230 &\\
            & SW,PA230 &\\
3C 303.1    & nuc & \multicolumn{1}{l}{$<$16} &
\multicolumn{1}{l}{15$^{+2}_{-1}$} \\
PKS 0023-26 & nuc  & \multicolumn{1}{l}{22$^{+5}_{-3}$} &
\multicolumn{1}{l}{$<$35}\\
PKS 0252-71 & nuc  &32$^{+20}_{-10}$/30$^{+80}_{-10}$  &110$^{+100}_{-60}$&&&\\
PKS 1306-09 & nuc  &90$^{+40}_{-90}$/40$^{+40}_{-10}$\\
PKS 1549-79 & nuc  & \multicolumn{1}{l}{14$^{+4}_{2}$}\\
            & PA75    & \\
            & PA-5    & \\
PKS 1814-63 & nuc  & \\
            & ext & \\
PKS 1934-63 & nuc  & \\
PKS 2135-20 & nuc  & 10$^{+1}_{-1}$&   $<$18 & \\
3C 459      & nuc  & \multicolumn{1}{l}{$<$46} & &
\multicolumn{1}{l}{$<$66}\\
            & ext & \\
\hline\hline
\end{tabular}
}
\end{table*}

\subsection{Reddening}
\label{sect:reddening}
\begin{figure}
\centerline{\psfig{file=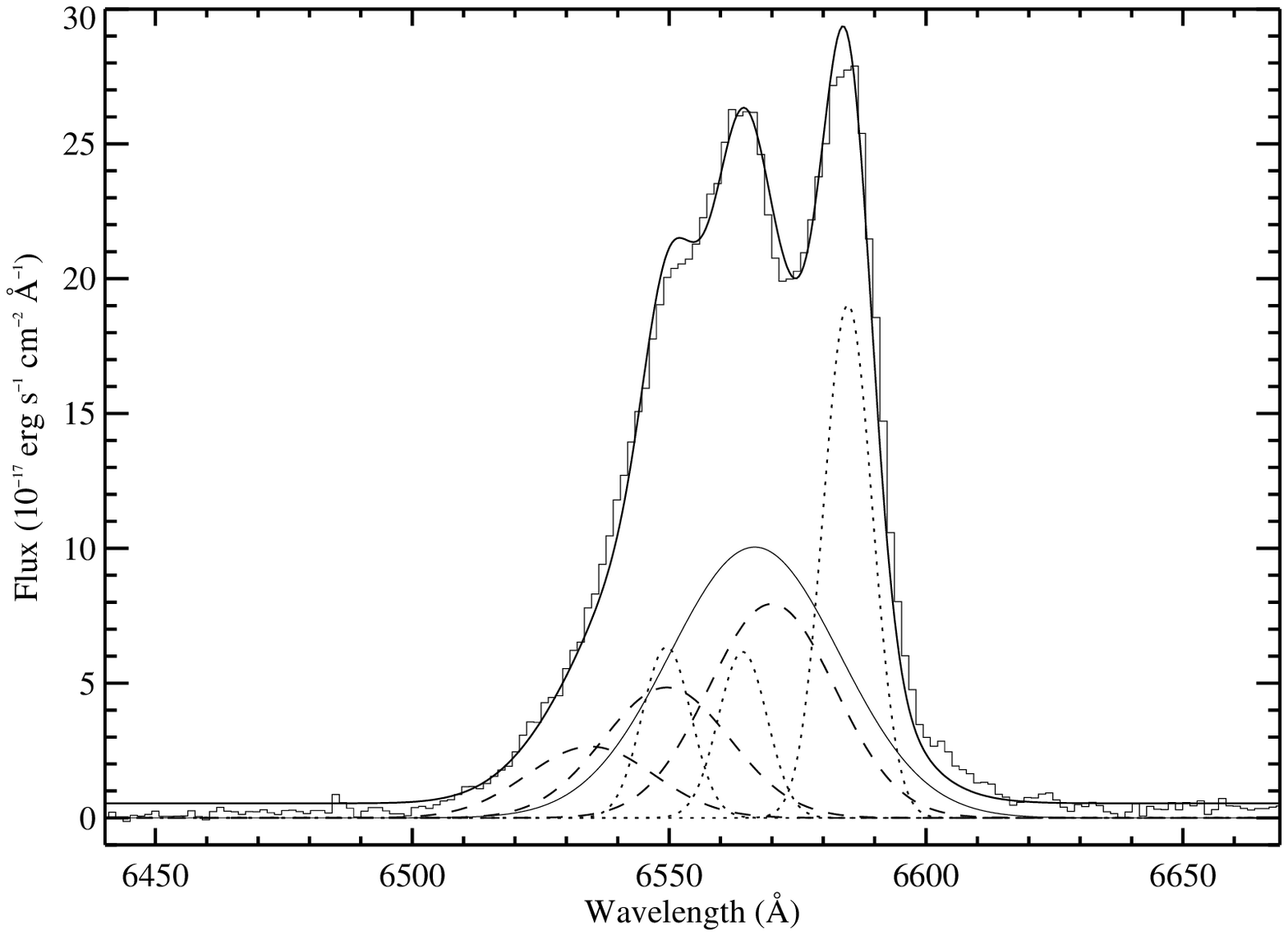,width=9cm,angle=0.}}
\caption[Example of H$\alpha$,{[N II]} blend]{The H$\alpha$,{[N
  II]}\lala6548,6584 blend in
  the nuclear aperture of PKS 1549-79. The faint solid line (histogram) is the
  data, the bold line is the overall best fitting model to the
  doublet and the dotted and dashed lines are the narrow and
  intermediate components (3 of each: 1 from H$\alpha$ and 2 for {[N
      II]}). The solid Gaussian is the broad component detected only
  in H$\alpha$ in the optical (see H06 for details). This plot
  highlights the complexity of the H$\alpha$/{[N II]} blend and the
  difficulties in measuring the reddening using H$\alpha$/H$\beta$
  ratio. }
\label{fig:han2}
\end{figure}
In Section \ref{sect:density}, and H03, we showed that there is a
suggestion that the  nuclear regions of our sample of compact radio
sources may harbour a dense ISM ($n_{e}$ 
$>$ few 1000 
cm$^{-3}$). Observationally, in addition to high densities, large
amounts of gas and dust can also be 
detected as reddening of the optical spectrum.

We have investigated reddening in all apertures in our sample using
the Balmer line ratios (e.g. \ha/\hb, H$\gamma$/\hb~and
H$\delta$/\hb), assuming a simple foreground screen model for 
interstellar dust \citep{seaton79}. The corresponding \ebv~values for each
object/aperture are summarised in Table \ref{tab:physcond}. 

Note, as discussed in Section 2 and H08, many galaxies in the sample
exhibit stellar 
absorption features. Whilst the continuum modelling and subtraction
attempts to remove these features, if a weak Balmer emission line is
coincident with a strong absorption feature (e.g. H$\delta$ and, to a
lesser extent, H$\gamma$), this may result in an incorrect (large) estimate of
the reddening 
\citep{fosbury01}. When modelling and subtracting the continuum (see
H08), the effect of the underlying absorption lines on the emission
lines was investigated, focussing in particular on the different
components of \hb. The well defined narrow components of the emission
lines are often relatively unaffected by the subtraction of different
models. However, the flux in the broader components could be affected
by as much as 10\% or more. Hence, wherever possible, estimates of
reddening from the stronger Balmer lines are used. It should also be
noted that in the more extreme sources, with several components to the
emission lines, there is also difficulty in accurately modelling
\ha, due to the complexity of the blend with the {[N II]}
doublet. Figure \ref{fig:han2} demonstrates this for the nuclear
aperture of PKS 1549-79.

The degree of estimated reddening  varies over a
  wide range, from negligible through to highly extinguished. There
  does not appear to be any particular trend with radio source size
  (i.e. GPS or CSS) although some of the highest  \ebv~values
  are measured in the GPS sources (e.g. \ebv~= 1.0 in 4C 32.44 and
  \ebv~= 1.4 in PKS 1345+12).   The main trend appears to be the
  increase in reddening with the FWHM of the emission line component
  {\it within} a particular object. In 7/14 sources (denoted by the
  stars in Table \ref{tab:physcond}), the
  degree of reddening increases significantly from the narrow
  component, often suffering negligible reddening (although some
  objects exhibit narrow components with moderate reddening), through
  to the intermediate and broad components where reddening is almost
  always significant. This effect was first observed in the GPS source
  PKS 1345+12 and is discussed by H03. Of the remaining 7
  objects which do not show this effect, it is not possible to do this
  test in 2 sources (i.e. both components observed have similar FWHM as in PKS
  1306-09) and in 1 further object, PKS 1814-63, the degree of
  reddening is not estimated. Taking this into account, 7/11 sources
  (or 67 \%) show the trend of increasing reddening with increasing
  FWHM. Hence, all reddening estimates are consistent with  dense and
  dusty cocoons surrounding the young radio sources, and in more than
  half of the sources observed there is strong evidence for
  stratification of the ISM as in PKS 1345+12. Such strong reddening
  in the nuclear regions of radio sources is not uncommon and is
  observed in a number of extended radio sources such as  Cygnus A
  (\citealt{tadhunter94,taylor03}).

Whilst we can see the general trends in the reddening and draw
conclusions about the existence of a dense and dusty circumnuclear
cocoon, given the various uncertainties discussed above, we have not
corrected the line fluxes for reddening. For some line ratios
(e.g. {[S II]} for the density), the lines are close together and
reddening effects will be negligible. For the diagnostic diagrams
presented later, for ratios in which the lines have a large separation
 in wavelength, we plot the uncorrected values.

\subsection{Electron temperature}
\label{sect:temp}
The electron temperature can be an important diagnostic for
distinguishing between the different ionisation mechanisms. For
example, one would expect to observe higher temperatures if the gas is
shock ionised rather than  photoionised by the AGN. 

The electron temperature, T$_{e}$, can be derived from the temperature
diagnostic {[O III]}(4959 + 5007)/4363. Where the weaker {[O III]}$\lambda$4363
line is detected, we have calculated the electron temperature, assuming
an electron density of 10$^{3}$ cm$^{-2}$. The derived electron
temperatures are summarised in Table \ref{tab:physcond}. 

No clear trends are observed across the
  entire sample but {\it all} measured temperatures are relatively high (in the
  main $\gtrsim$ 14,000 K), and in some objects (e.g. 4C 32.44) a
  gradient with FWHM is   observed, although the uncertainties are large. 
Such   temperatures are too high to be accounted for by simple single slab
  photoionisation models which generally predict temperatures of
  $\lesssim$ 11,000 K \citep{tadhunter89} and may be a signature of shock
  ionisation, although mixed-medium ionisation models could also account for
  such high temperatures \citep{binette96}. Density effects may also
  be responsible for the unrealistically high temperatures measured
  in  some sources/apertures.

\section{Discussion}

\subsection{Do compact radio sources harbour a dense and dusty natal
  coccoon?}

\citet{tadhunter01} concluded that the compact, flat spectrum
radio source PKS 1549-79 is a young, recently triggered radio-loud
AGN which is still surrounded by a dense and dusty natal coccoon,
obscuring the active nucleus from view. This was confirmed by the
detection of the proto-quasar, initially at near infra-red wavelengths
\citep{bellamy03} and, later, at the reddest optical wavelengths in
higher quality optical spectra (H06)
Deep optical images also reveal that PKS 1549-79 has recently been involved in
a major merger event. Similarly, H03 
reported the discovery of significant reddening and high
electron densities in the nuclear aperture of the GPS source PKS
1345+12. Again these results were interpreted as the signature of a
dense and dusty natal cocoon, particularly as PKS 1345+12 also shows
evidence of a recent merger (double nucleus in a common envelope
e.g. \citealt{axon00}). 

In Sections {\ref{sect:density}} and {\ref{sect:reddening}},
we investigated the density and reddening in the nuclear apertures of
14 compact radio sources, including PKS 1345+12 and PKS
1549-79. Whilst it has been difficult to obtain unique results,
particularly for the electron densities, we can draw some general
conclusions. The majority of the sources in the sample do tend to show
evidence for high densities and large reddening in the nuclear
regions, particularly in the broader shifted components. This trend is
particularly apparent in the reddening  -- 7/14 sources show
convincing evidence for increasing reddening with FWHM (2 GPS, 3 CSS
\& 2 compact core sources). 

Hence, it is clear that  the nuclear regions of compact radio sources 
contain large amounts of gas and dust. This is consistent with the
large reddening observed in the optical spectra of CSS quasars 
\citep{baker95}, although at X-ray wavelengths, the evidence for
differences in column densities between young AGN and more extended
radio sources is contradictory. \citet{guainazzi06} report high
column densities in GPS sources, for all
sources in their small sample 
with good quality hard x-ray data. The measured column densities in
GPS sources are higher   compared to FR I and BLR FR II sources, but
comparable to high-ionization FR II sources, and so they  suggest the
obscuring material is located in an obscuring torus. However,
\citet{vink06} find a variety of column densities in GPS and CSS
sources, with no evidence for higher column densities compared to more
extended radio sources. This latter observation is consistent with the
fact that strong optical reddening is not uncommon in the nuclear regions
 of extended radio sources
(e.g. \citealt{robinson87,tadhunter94b,robinson00,robinson01,taylor03}).

Whilst there is evidence for a dense and dusty circumnuclear cocoon,
there appears to be insufficient material to constrain and frustrate
the radio sources (e.g. PKS 1345+12: \citealt{holt03}). The only way
frustration would be possible is if the matter were distributed in
giant clouds, which may be able to disrupt a radio jet
\citep{bicknell03}. Hence, the observations are consistent with compact
radio sources being young, recently triggered radio sources residing
in dense and dusty natal cocoons, rather than old, confined and
frustrated radio sources.

\subsection{What is the dominant ionisation mechanism for the gas?}

\begin{figure*}
\centerline{\psfig{file=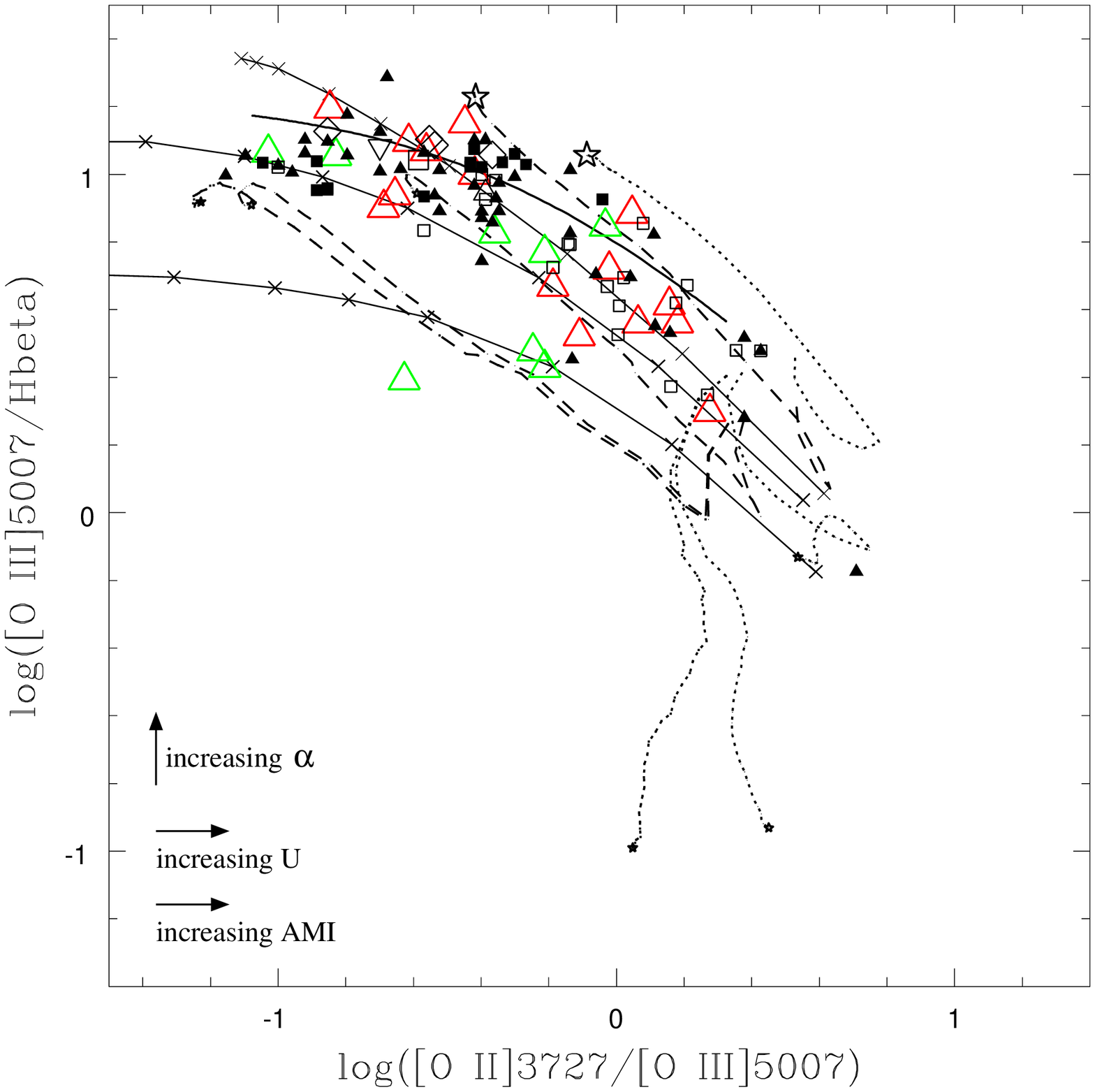,width=14.0cm,angle=0.}} 
\centerline{ (a)}
\caption[Diagnostic diagrams for the sample: nuclear narrow
  components]{Diagnostic diagrams for the entire sample of nuclear
  narrow components (large 
  red open triangles) and extended components (large green open
  triangles). Arrows represent upper limits. The ionisation
  models plotted are as follows: 
  \newline 
-- crosses joined by solid black lines: optically thick power-law
(AGN) photoionisation sequence in the ionisation parameter $U$ (2.5
$\times$ 10$^{-3}$ $<$ $U$ $<$ 10$^{-1}$)  from 
 {\sc mappings} for $\alpha$=-1.0, -1.5 \&
-2.0; \newline
-- bold black solid line: mixed-medium models (including both
matter-bounded and ionisation-bounded clouds; 10$^{-2}$ $\leq$
A$_{M/I}$ $\leq$ 10) from \protect\citet{binette96}; \newline
--  dotted lines: new shock models from
\protect\citet{allen08}. Sequences are for magnetic paramter,
B/$\sqrt{n}$ $=$ 
0.01, 1, 10, 100, 1000 $\mu$G cm$^{3/2}$, with shock velocities
10$^{2}$--10$^{4}$ \kms~and pre-shocked gas density n$_{e}$ = 100 cm$^{-3}$. The stars identify the
highest shock velocity (10$^{4}$ \kms) on each track with the large
open star identifying the track with the highest magnetic
paramter; \newline
-- dashed lines:  new shock plus precursor models from
\protect\citet{allen08}. Sequences are for magnetic paramter, B/$\sqrt{n}$ $=$
0.01, 1, 10, 100, 1000 $\mu$G cm$^{3/2}$ with shock velocities
10$^{2}$--10$^{4}$ \kms~and pre-shocked gas density n$_{e}$ = 100
cm$^{-3}$.  The stars identify the
highest shock velocity (10$^{4}$ \kms) on each track with the large
open star identifying the track with the highest magnetic
paramter. \newline
For comparison, various data from the literature are overplotted (small
black points): nuclear regions (filled triangles: 
\protect\citealt{wills02,storchi96,tadhunter87,robinson87},
\protect\citealt{grandi78,costero77,cohen81,dickson97},
\protect\citealt{carmen01,villar98,clark98,robinson00});  
  extended emission line regions (EELRs, filled squares:
  \protect\citealt{tadhunter94,morganti91,storchi96,robinson00});
    EELRs with evidence for jet-cloud interactions (open squares:
    \protect\citet{villar98,clark98,carmen01}) and all regions
   in Cygnus A (larger black open symbols: Taylor et al. 2003).
}
\label{fig:diagsamplen}
\end{figure*}
\setcounter{figure}{2}
\begin{figure*}
\centerline{\psfig{file=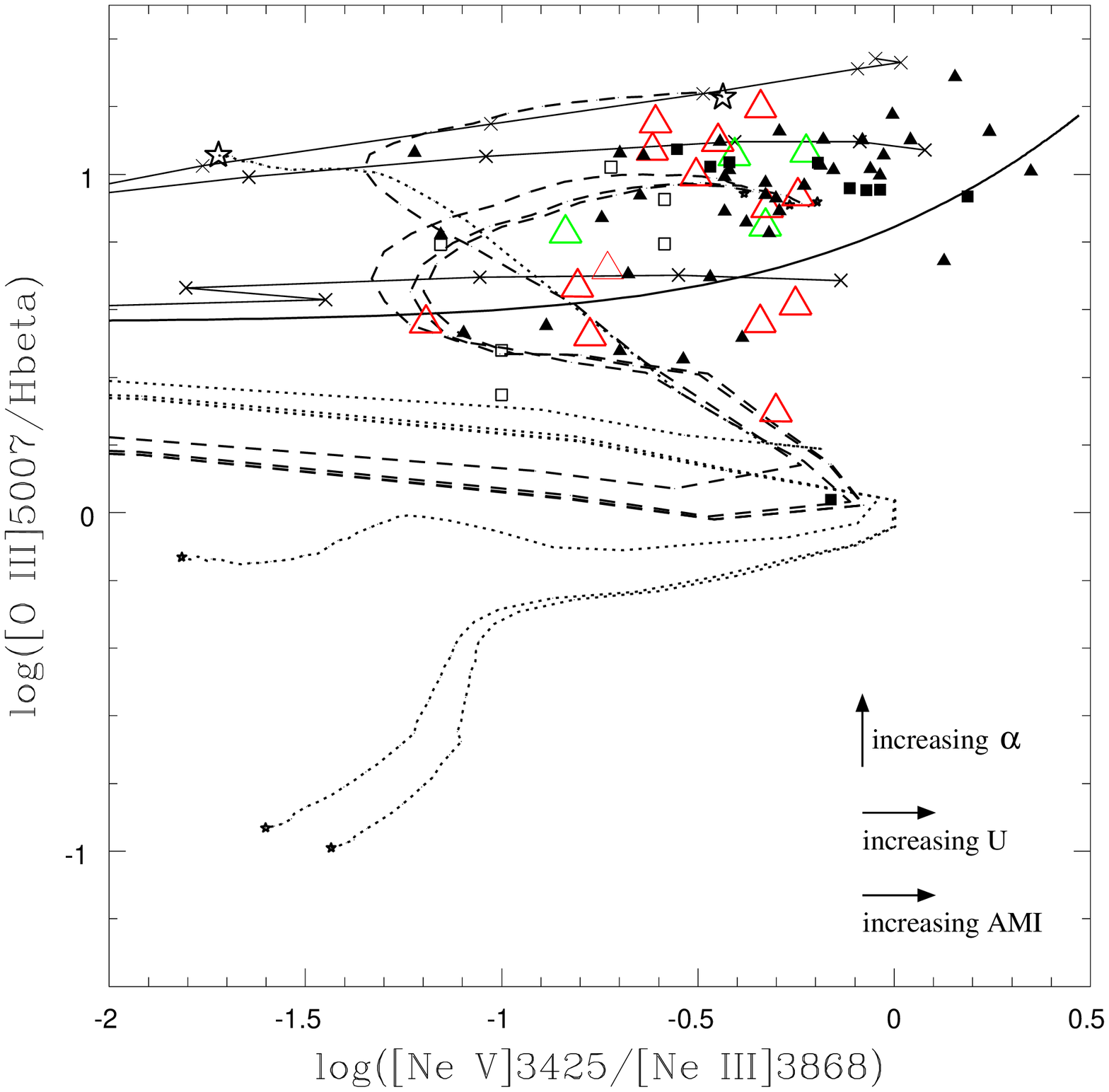,width=11.5cm,angle=0.}}
\centerline{ (b)}
\centerline{\psfig{file=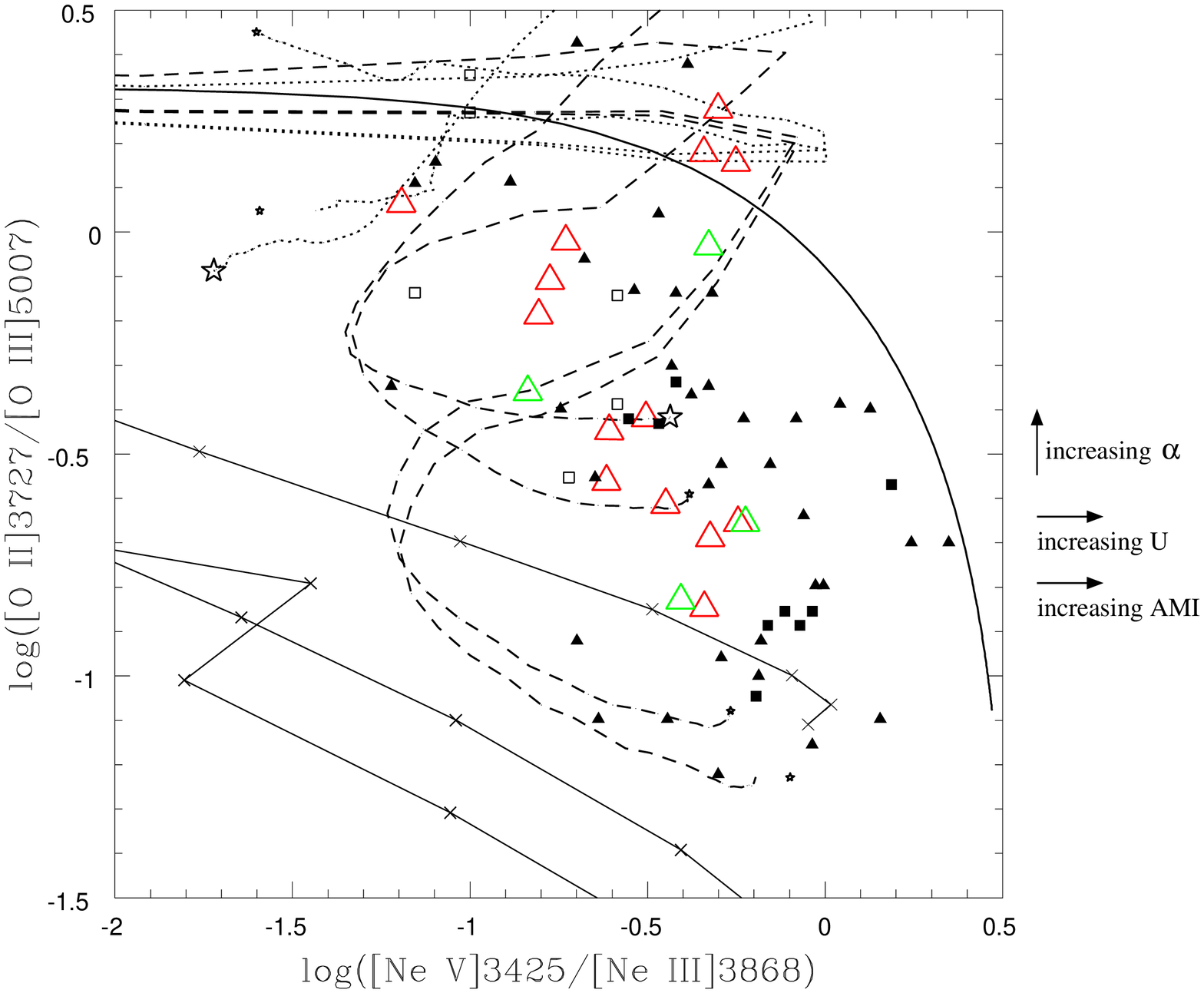,width=13.0cm,angle=0.}}
\centerline{ (c)}\caption[Nuclear narrow components]{\it continued.}
\end{figure*}
\setcounter{figure}{2}
\begin{figure*}
\centerline{\psfig{file=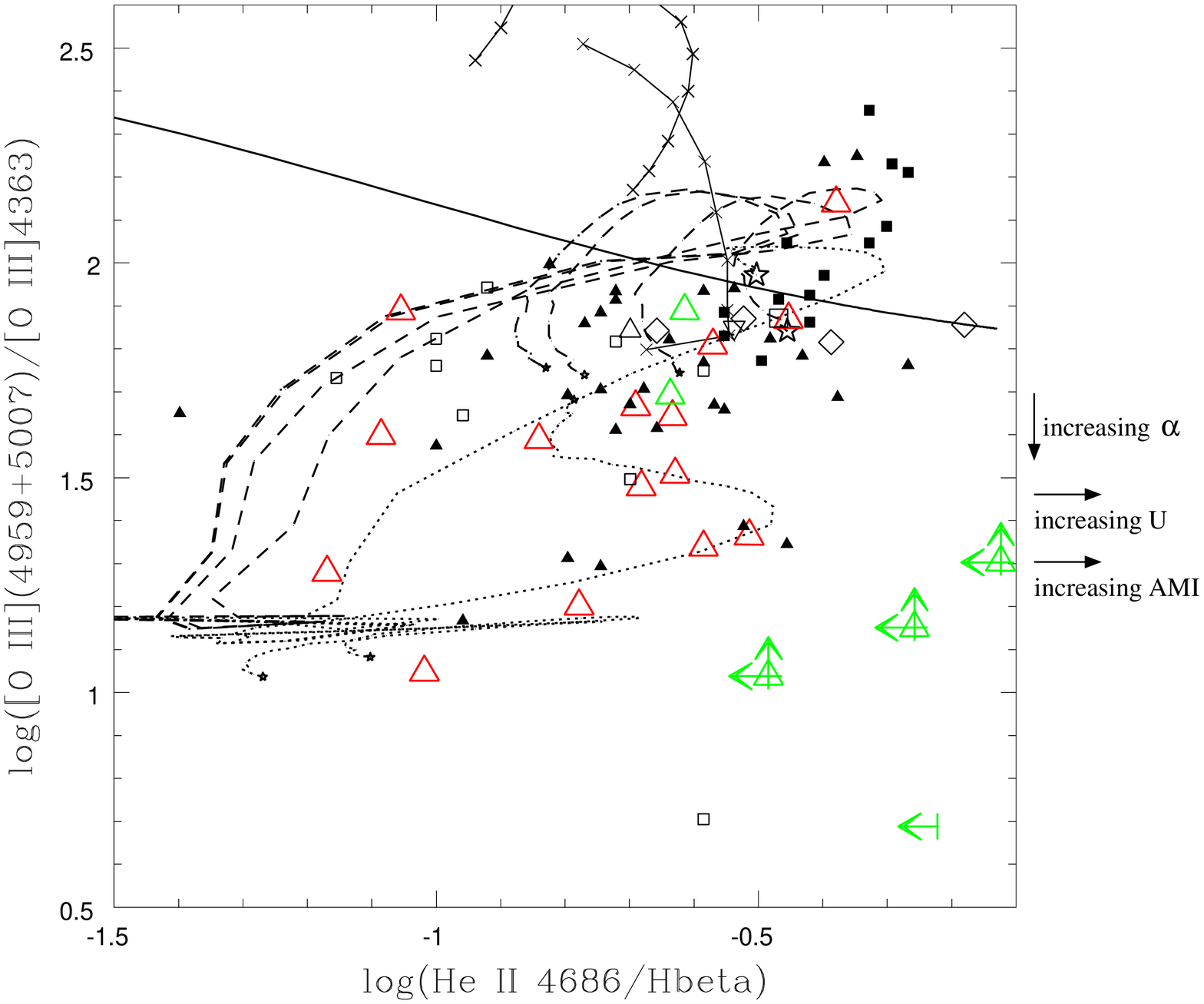,width=13.0cm,angle=0.} }
\centerline{ (d)}
\centerline{\psfig{file=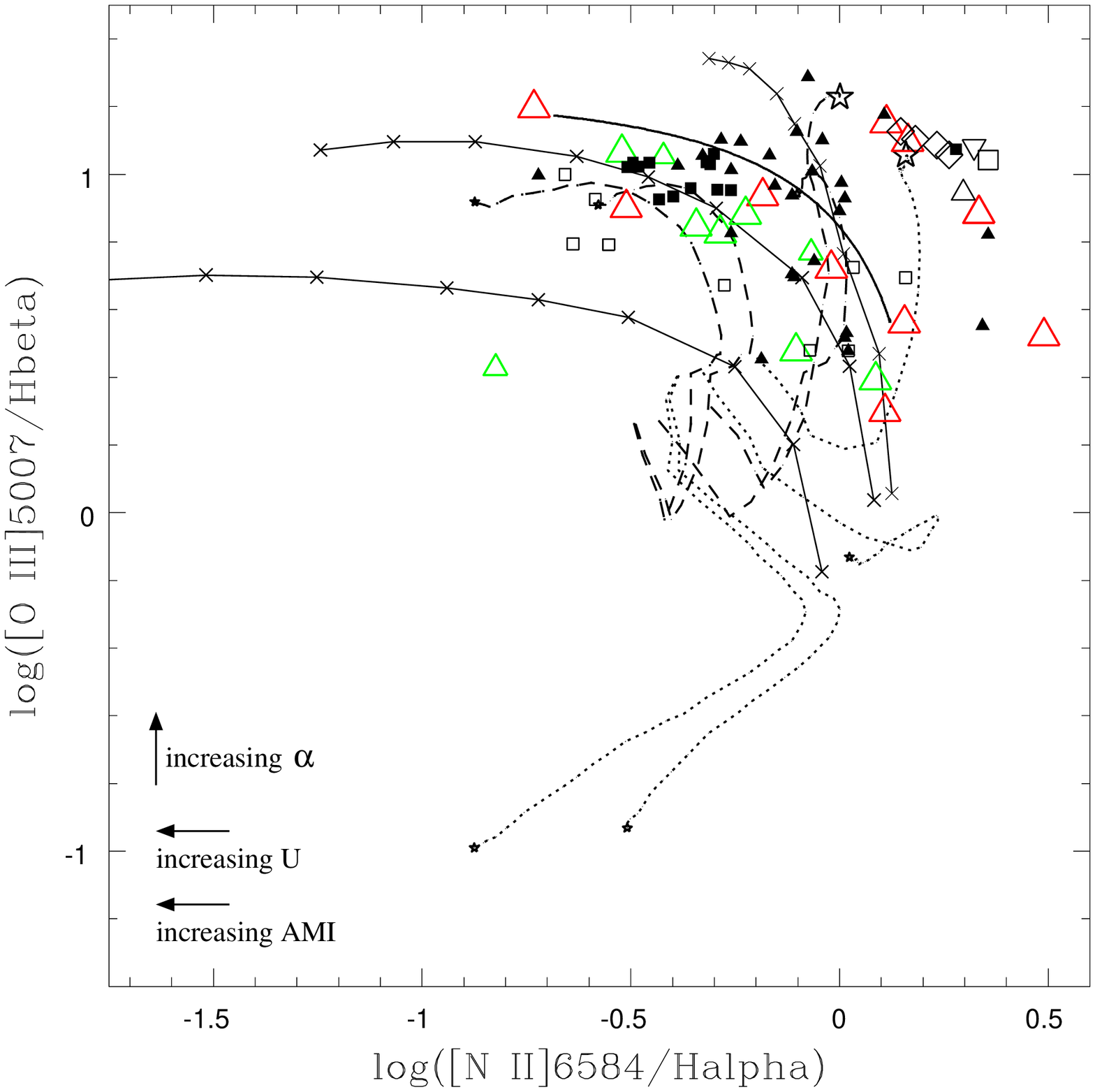,width=11.5cm,angle=0.} }
\centerline{ (e)}\caption[Nuclear narrow components]{\it continued.}
\end{figure*}
\vspace*{-0.3cm}
\begin{figure*}
\vspace*{-0.3cm}\centerline{\psfig{file=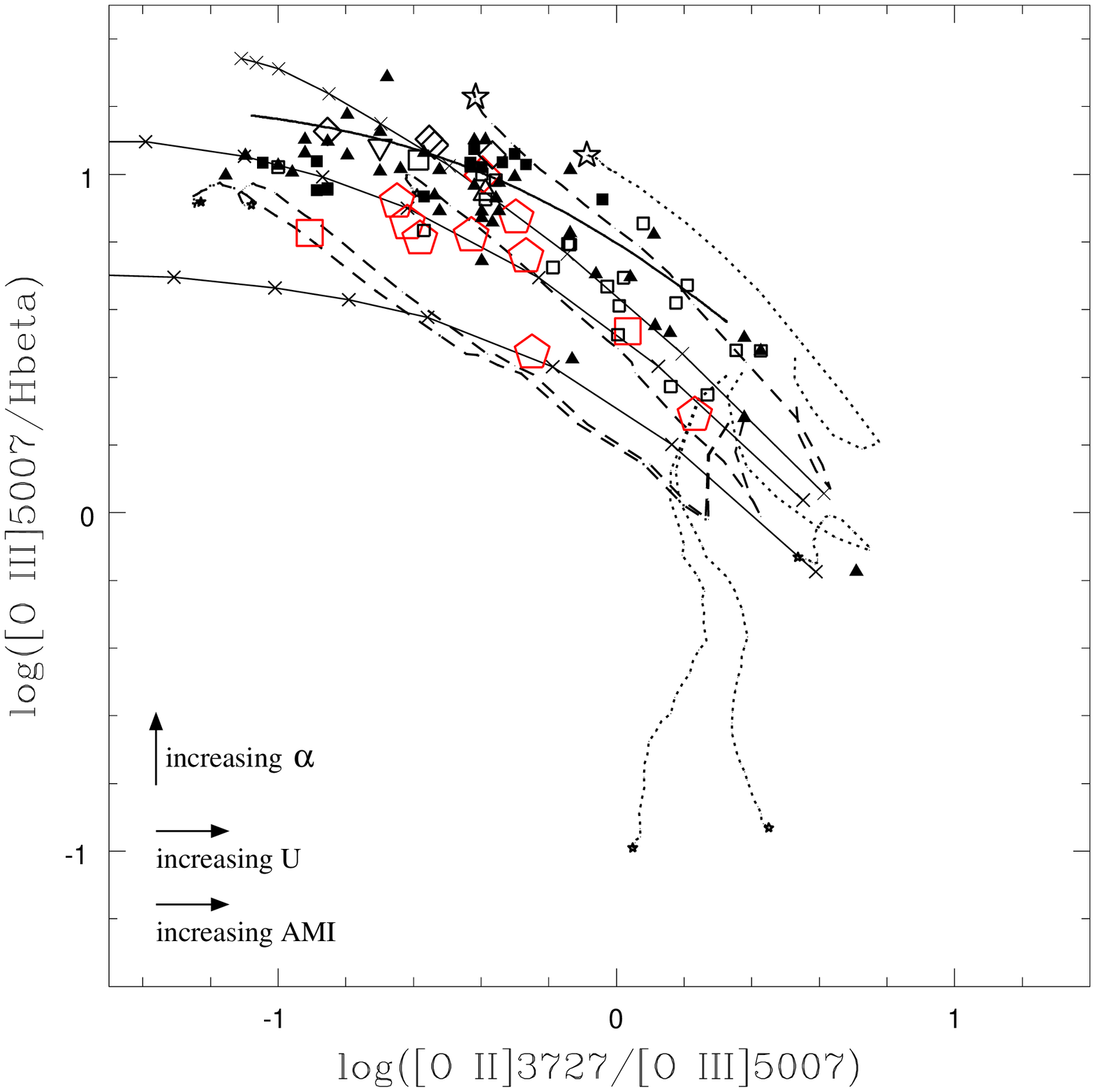,width=11.5cm,angle=0.} }
\centerline{ (a)}
\centerline{\psfig{file=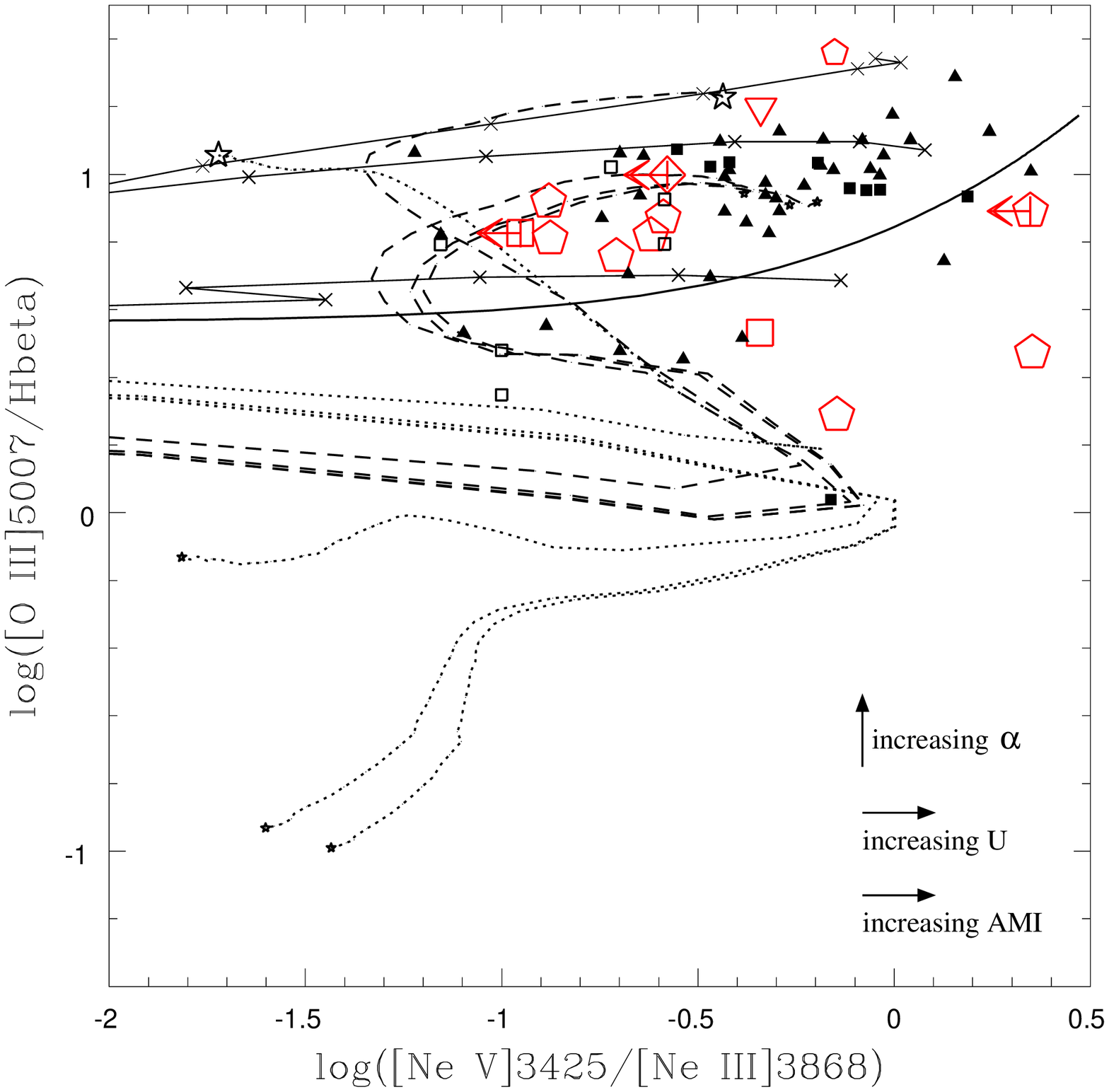,width=11.5cm,angle=0.} }
\centerline{ (b)}\caption[Diagnostic diagrams for the sample: nuclear shifted (intermediate, broad and very broadcomponents).]
{Same as for Figure \protect\ref{fig:diagsamplen} but 
for the shifted components: intermediate (red open pentagons), broad
(red large open squares) and very broad (red large open
diamonds). Small black points represent the comparison sample (see
caption of Figure \protect\ref{fig:diagsamplen} for details).}
\label{fig:diagsampleb}
\end{figure*}
\setcounter{figure}{3}
\begin{figure*}
\centerline{\psfig{file=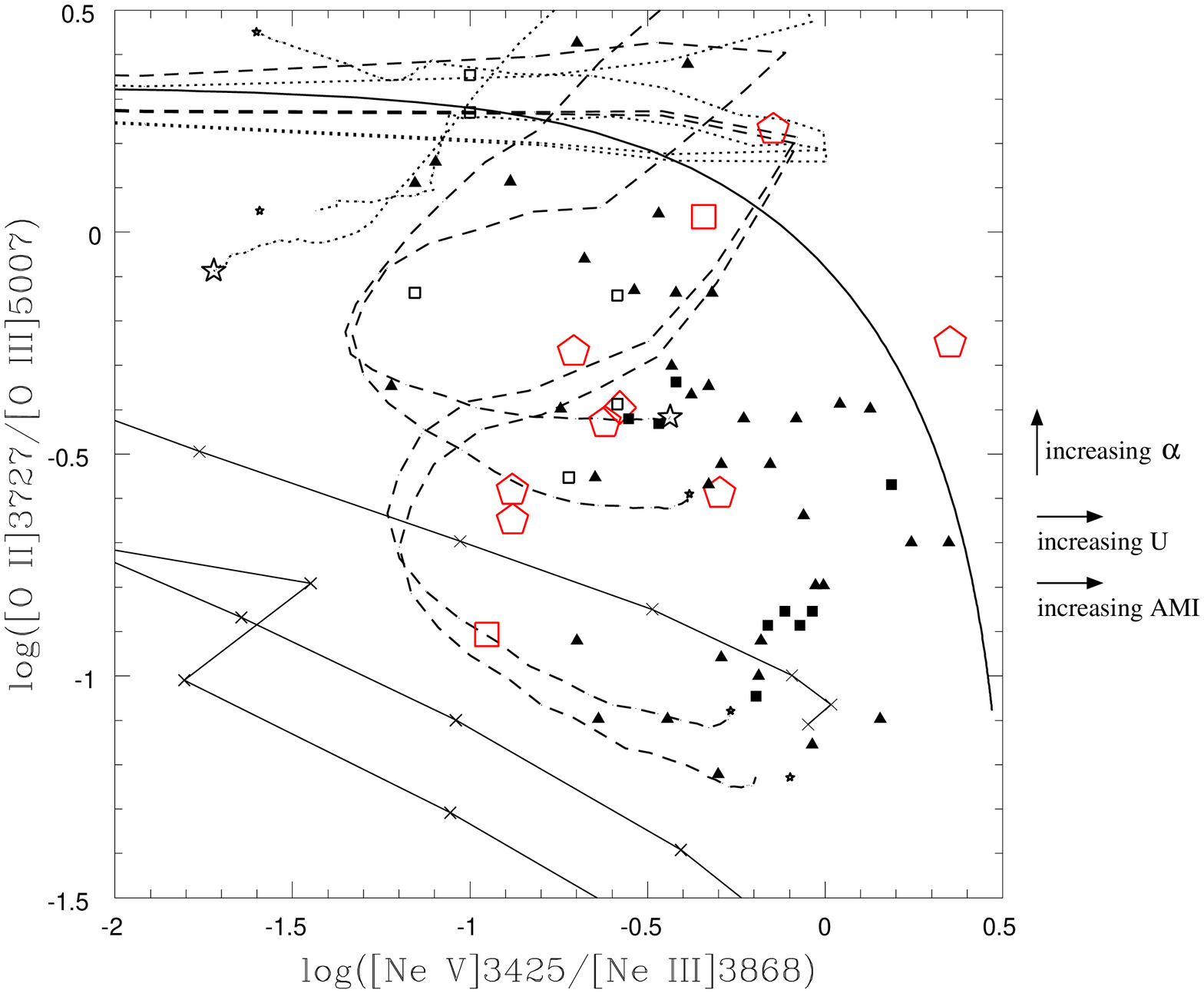,width=13.0cm,angle=0.} }
\centerline{ (c)}
\centerline{\psfig{file=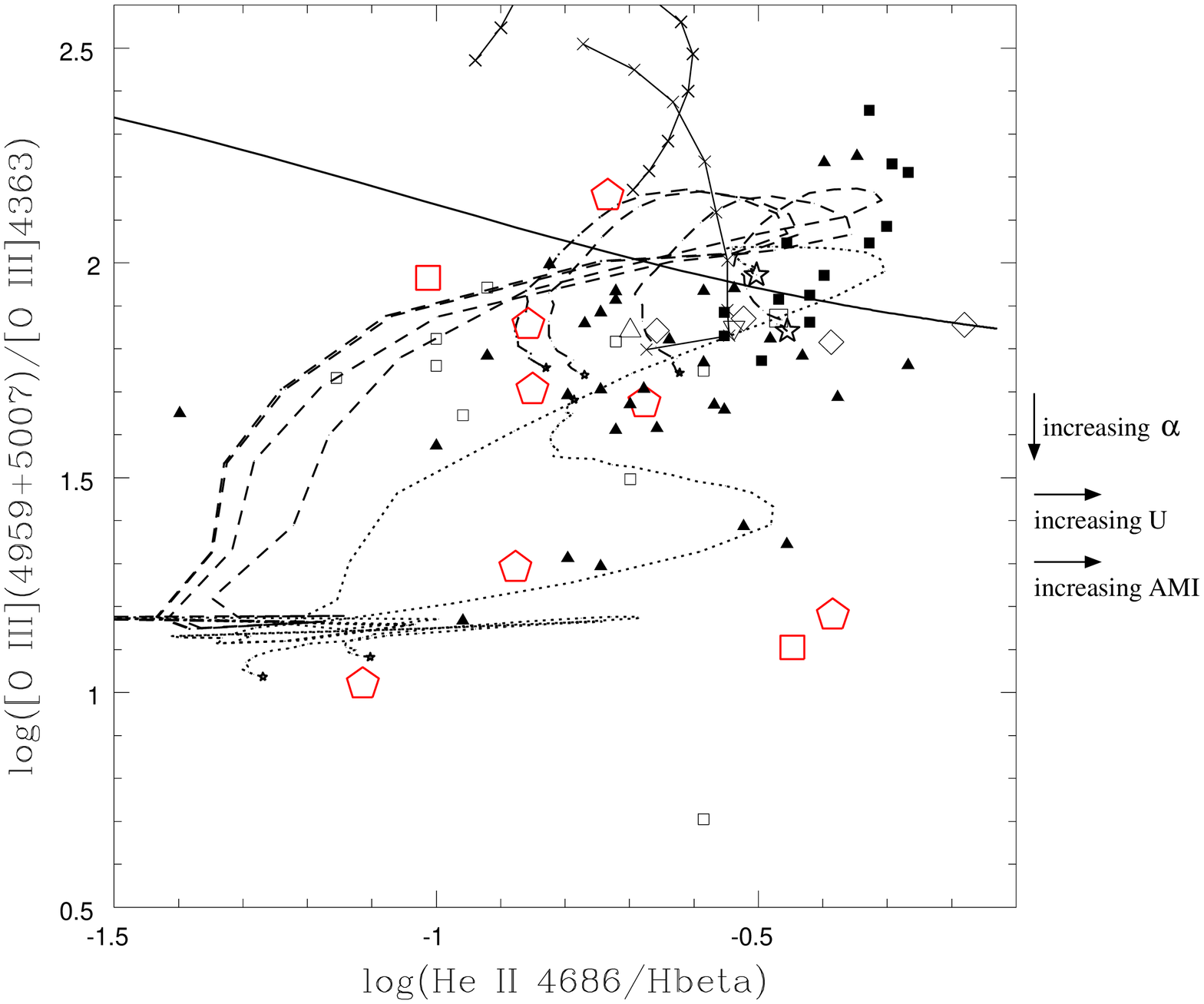,width=13.0cm,angle=0.}} 
\centerline{ (d)}\caption[Nuclear shifted components]{\it continued.}
\end{figure*}
\setcounter{figure}{3}
\begin{figure*}
\centerline{\psfig{file=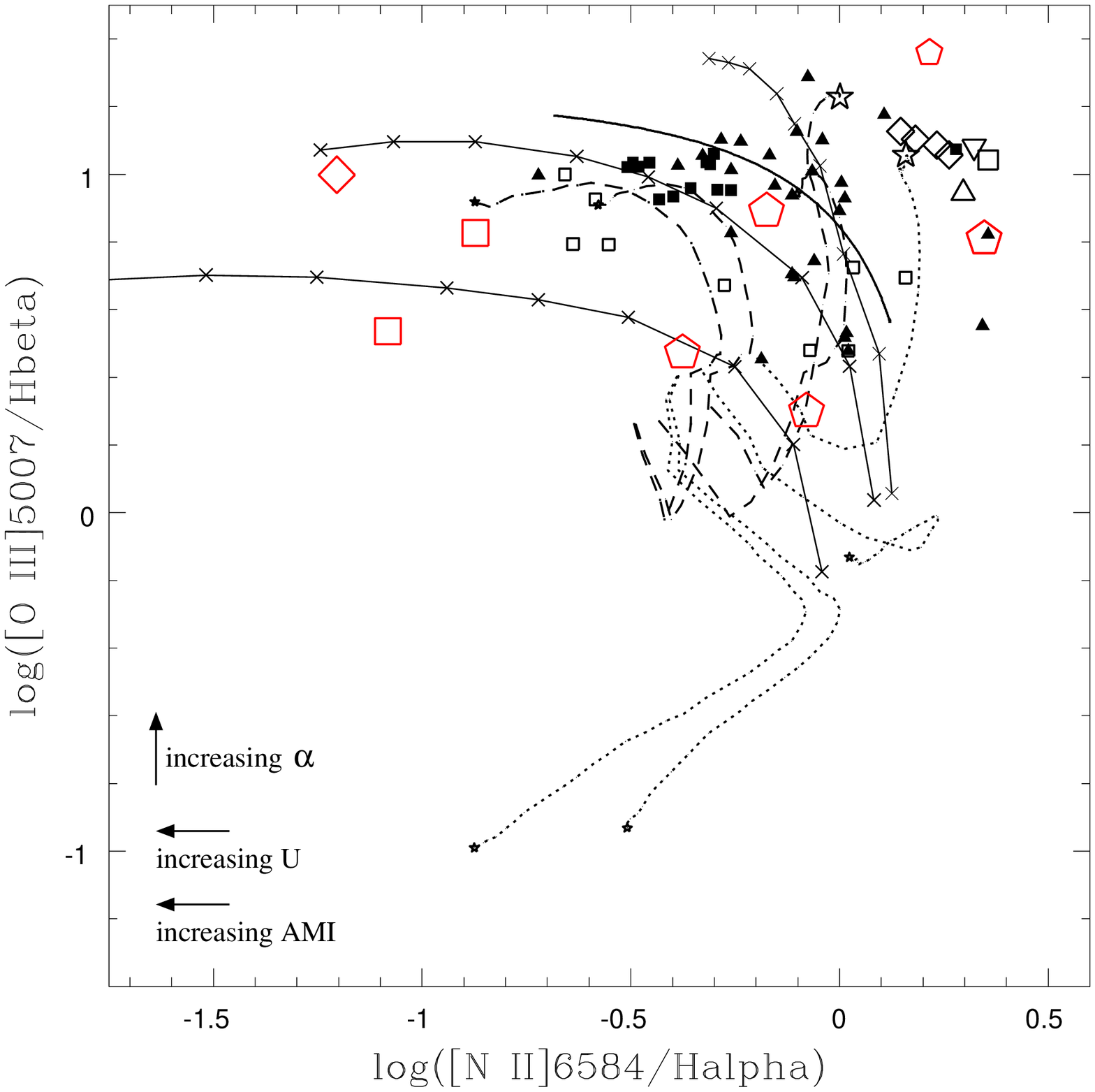,width=11.5cm,angle=0.}}
\centerline{ (e)}\caption[Nuclear shifted components]{\it continued.}
\end{figure*}
\vglue 0.2cm\noindent

\citet{tadhunter01} suggested  that the outflow in the
compact radio source PKS 1549-79 was driven by the expanding radio
jets. Certainly, jet-driven outflows are a convenient way of
explaining the extreme kinematics observed in the optical emission
line gas. However, it is also possible that the outflows have been
accelerated by some other mechanism, for example, quasar-induced winds
or starburst driven superwinds.

A key way to try to distinguish between these scenarios is the use of
emission line ratios and diagnostic diagrams, a technique successfully
employed for identifying jet-cloud interactions in the extended emission
line regions of extended radio sources
(e.g. \citealt{villar98,clark98,carmen01}). We have therefore plotted a
variety of diagnostic diagrams for the different emission line
components in Figure \ref{fig:diagsamplen} (nuclear narrow \&
extended components) and Figure \ref{fig:diagsampleb} (nuclear
broader/shifted components). In addition to the data for this sample,
we have also plotted a comparison sample of nearby extended radio
galaxies with detailed line ratio information (see the caption of
Figure 3 for details)  and model results for simple slab AGN
photoionisation (from MAPPINGS Ic: see \citealt{ferruit97}), mixed-medium photoionisation
\citep{binette96} and the latest shock and shock plus precursor models
from \citet{allen08}. Full details of the models are given in the
caption of Figure \ref{fig:diagsamplen}. Note that it can be difficult
to distinguish between the AGN and shock plus precursor models for the
high shock velocity case, because the hard continuum generated in
the hot post-shock gas (that photoionizes the precursor) can mimic that
of an AGN nucleus. Moreover, the mixed medium models of \citet{binette96},
while successfully explaining the relatively strong high ionization
lines and high electron temperatures measured in some AGN, cover only a
limited range of parameter space for the matter- and radiation-bounded components
included in the models; better agreement with the observed line ratios
in a given object might be obtained by tuning the
input parameters of such models.

Initial inspection of Figures \ref{fig:diagsamplen} and
\ref{fig:diagsampleb} reveals that the positions of the broader and
narrower emission line
components of the compact radio sources are not significantly
different from each other, and are remarkably similar to those of the
comparison sample of extended radio sources. This latter result is
surprising given the striking differences in the nuclear emission line
kinematics and the strong radio-optical alignments reported for CSS
sources (see Section 4.3).

In terms of the results for the compact radio sources, whilst the
diagrams are complex, it is still possible to identify some
tendencies in the data when interpreting all diagrams together:
\begin{itemize}
\item {\bf Narrow components.} 
In the majority of the diagrams (e.g. Figures 3a, 3b, 3e), the nuclear
narrow components lie close 
to tracks for photoionisation (both simple AGN photoionsition and the
A$_{M/I}$ sequence) whilst in Figures 3c \& 3d, the nuclear narrow
components lie away from the plotted photoionisation tracks, and
appear more consistent with shock models. However, in the latter cases, the data always lie closer
 to the higher velocity end of the tracks (v$_{\rmn shock}$ $\geq$
500\kms). Given that the nuclear narrow components are relatively
quiescent, with small velocity 
widths (FWHM $\lesssim$ 400\kms), and no blueshifts (H08 demonstrated
that they are at the systemic velocity), such an interpretation would
be unlikely. It should also be noted that, in both Figures 3c \& 3d, the
locus of the data for the narrow components is similar to that for the
mixed-medium photoionisation models -- varying the parameters of the
mixed-medium models would result in tracks which  describe the
data better than the other models. We therefore interpret Figure 3 as
the nuclear narrow components 
showing a tendency to be consistent with AGN photoionisation (both
simple slab and mixed-medium models). The extended narrow components
are also plotted on Figure 3, and we see no significant differences
between the nuclear and extended narrow components suggesting they
have similar  ionisation mechanisms.

\item {\bf Broad, shifted components.} 
As for the nuclear narrow components, careful consideration of all information
together, along with details of the models, is required in order to 
disentangle the overlapping models on the  diagnostic
diagrams in Figure \ref{fig:diagsampleb}. From all the diagrams in Figure
4, we can rule out pure shock ionisation. In all 5 panels of Figure 4,
the data occupy the same region as the shock plus precursor models,
typically in the region of high velocity (v$_{\rmn shock}$ $\gtrsim$
500 \kms) shocks (this is most clearly seen in e.g. Figures 4a, 4b \&
4c). In some diagrams, the 
location can also be interpreted as consistency with AGN
photoionisation (e.g. Figures 4a \& 4b), but in Figures 4c \& 4d, the
data is neither in the same region as, nor has the same locus as, any of
the photoionisation models. Again, by
combining all diagrams, there appears to be a mild tendency towards shock
plus precursor models, with high shock velocities  (v$_{\rmn shock}$ $\gtrsim$
500 \kms) in the nuclear shifted components, although this evidence
is not conclusive. 

\item {\bf Object classification}. The {[O
    III]}$\lambda$5007/\hb~vs. {[N II]}$\lambda$6584/\ha~diagram
  (Figures 3e \& 4e for the narrow and broad components respectively) is a
    well known diagram to distinguish between the different classes of
    object and therefore the dominant ionisation mechanism
    (e.g. \citealt{kewley01,kauffmann03}). Almost all 
    components are consistent with AGNs/Seyferts, with a few in the
    region of the diagram occupied by LINERS. There are a
    couple of notable exceptions -- the extended region
    to the SE of the nucleus in 3C 277.1, and the extended regions to
    the SE and SW in PKS 1345+12 are all consistent with stellar
    photoionisation. The observation of ongoing star formation in
     radio galaxies is rare, with only a few other  cases known:
    e.g. Coma A \citep{solorzano03}; PKS 1932-46 \citep{villar05}. The
    Super Star Clusters (SSCs) in the halo of PKS 1345+12 are
    discussed in \citet{javi07}. 

\end{itemize}

\subsection{What drives the outflows?}

In H08 we reported evidence for extreme emission line outflows in the
nuclear regions of 11/14 compact radio sources. From the kinematics
alone, the large
velocity shifts and widths (FWHM) observed (up to 2000 \kms) are
entirely consistent with the idea that the young, expanding radio
source is strongly interacting with the natal cocoon and driving
outflows in the ISM, as suggested by the results for PKS 1549-79
\citep{tadhunter01}. This argument is further
strengthened by the growing evidence  in the literature that the line
emission in CSS sources is both on 
 similar scales to, and often closely aligned with, the 
radio source  \citep{devries97,devries98,axon00}. This effect is
observed for {\it all} CSS sources with HST imaging. Recent
HST imaging of two of the sources in our sample (PKS 1345+12 and PKS
1549-79) by \citet{batcheldor07} has also shown that in the smaller
GPS sources, the optical line emission is also on a similar scale to the  radio
source. However, although galaxy wide outflows
(i.e. starburst-driven winds) can be confidently ruled out for
GPS sources, on the
basis of emission line region morphology  it is not
possible to distinguish between AGN-winds and jet-cloud interactions,
which would both be expected on the nuclear scales\footnote{Due to the
  small spatial scales, it is currently not possible to resolve the
  outflowing regions sufficiently.}.  Whilst the kinematics and
radio-optical alignments are clearly suggestive of jet-cloud
interactions and radio jet-driven outflows, the above evidence alone
is, so far, circumstantial. 
\begin{figure*}
\begin{tabular}{cc}
(a) & (b)\\
\psfig{file=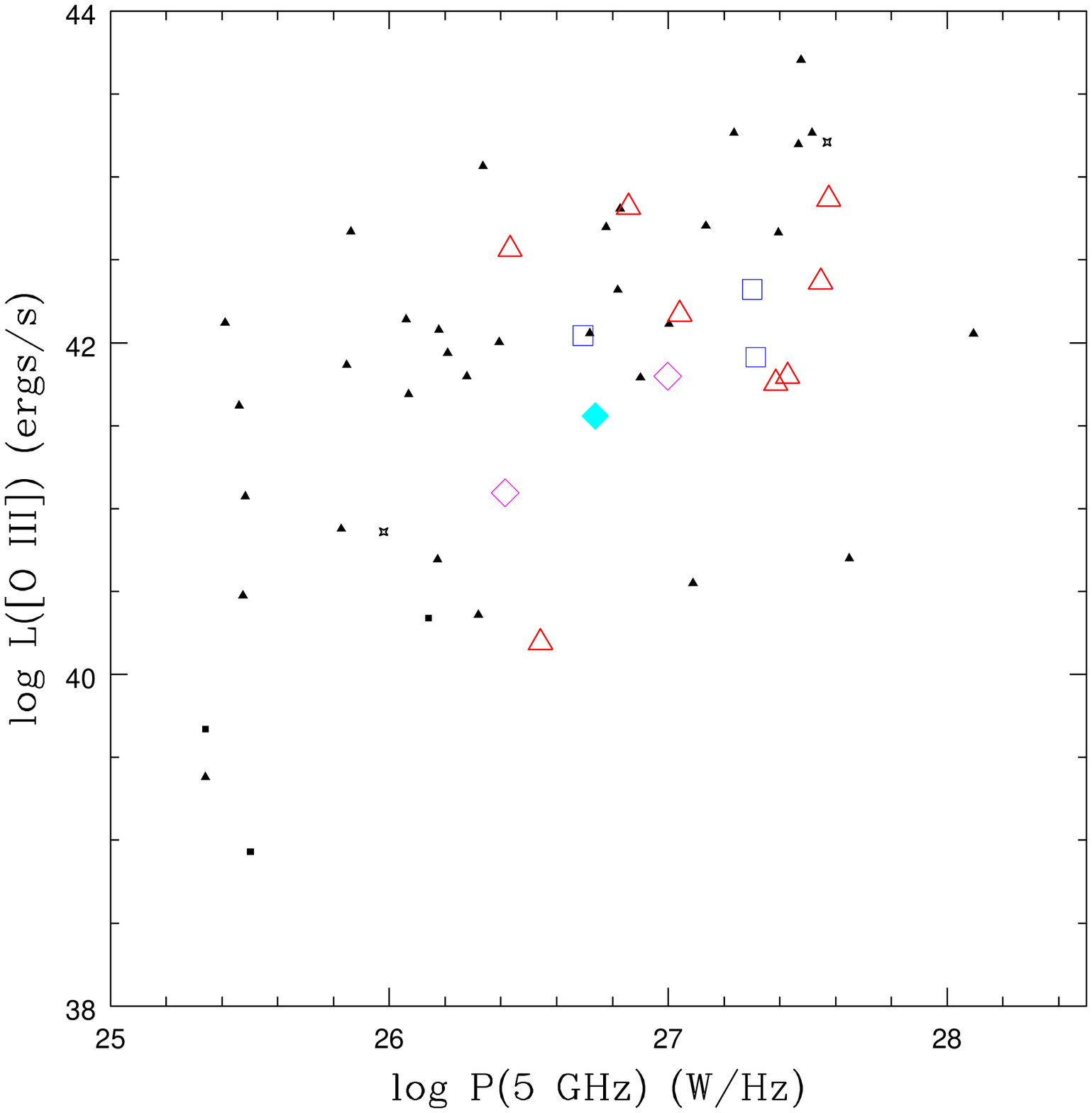,width=8cm,angle=0.}&
\psfig{file=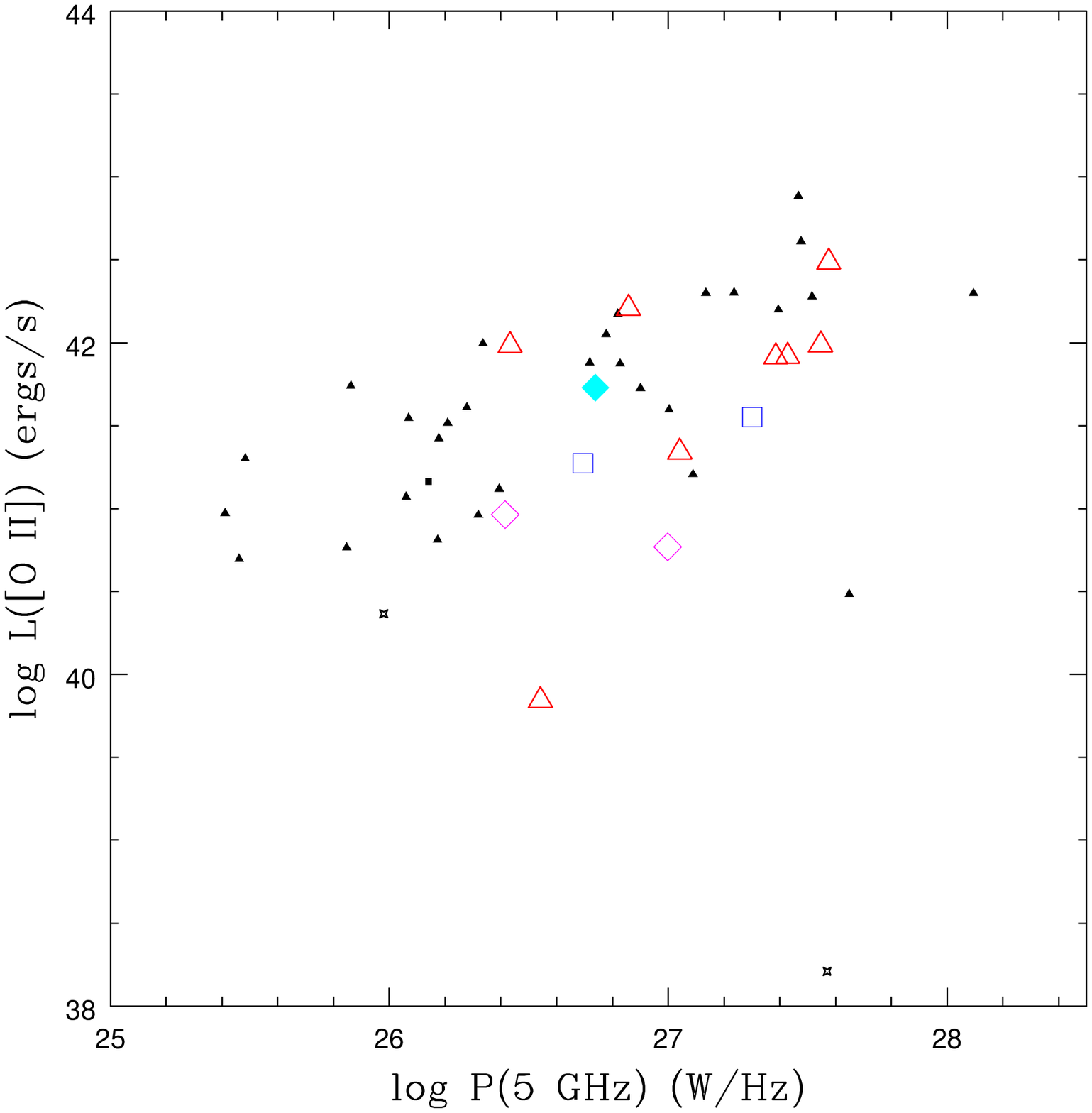,width=8cm,angle=0.}\\
\\
(c) & (d)\\
\psfig{file=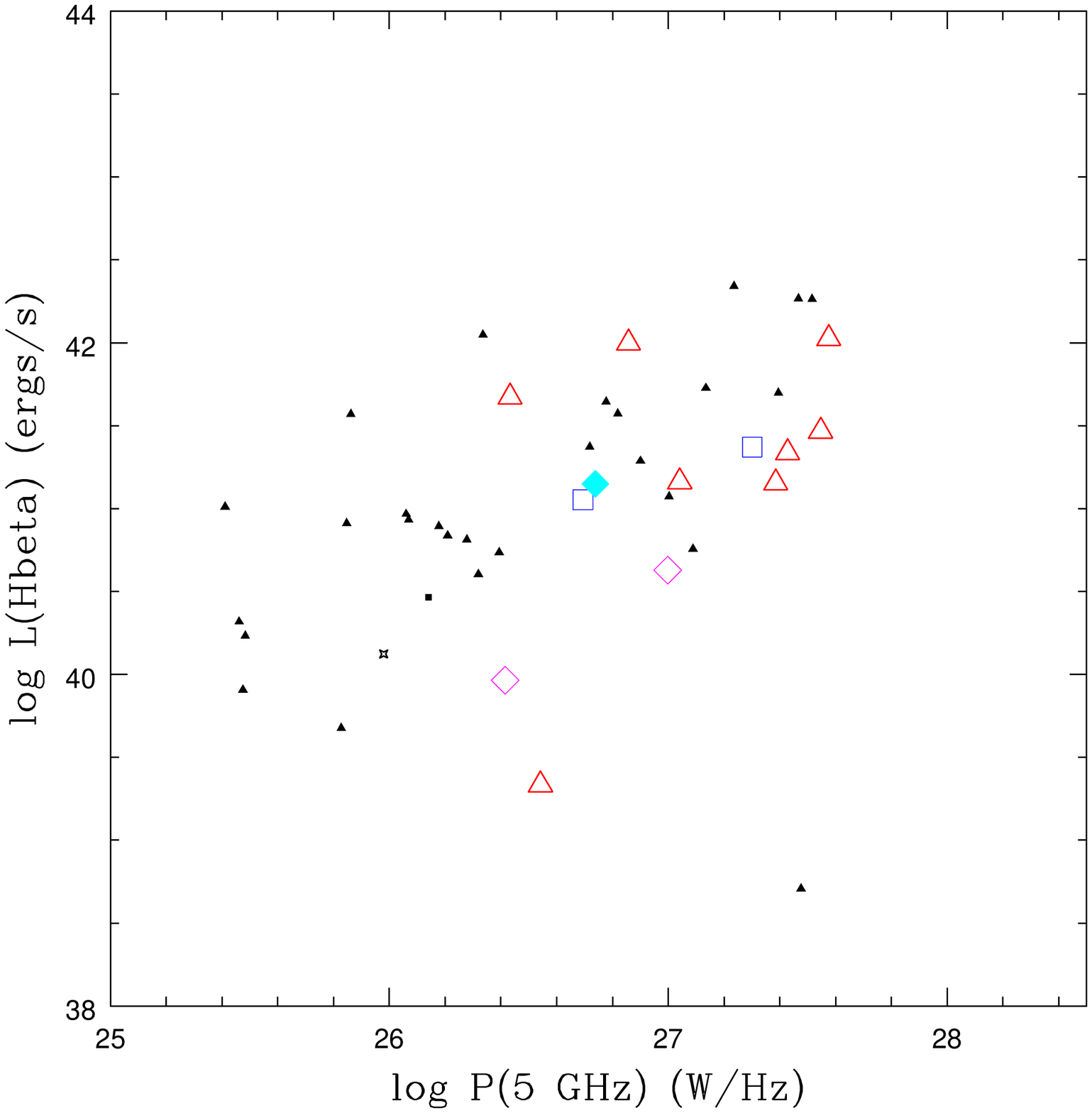,width=8cm,angle=0.}&
\psfig{file=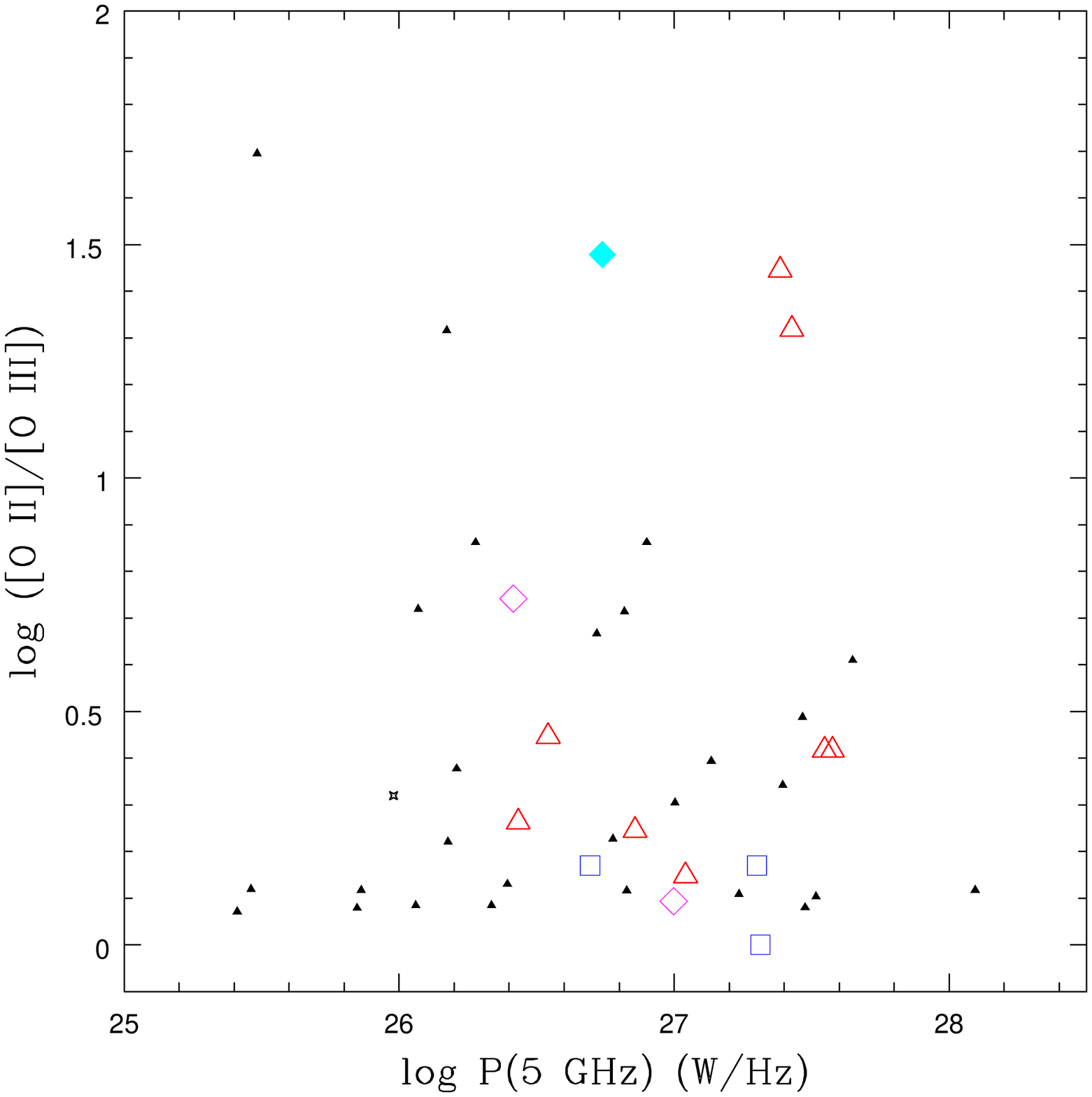,width=8cm,angle=0.}\\
\end{tabular}
\caption[Radio optical correlation plots]{Radio-optical correlation
plots. Plotted are the sources from this sample: CSS (red triangles),
GPS (blue squares), compact  flat
  spectrum (magenta diamonds) and compact core (cyan diamond). 
 The comparison sample is a redshift selected subsample (0.05 $<$ $z$
 $<$ 0.7) of the the 2 Jy
  sample (e.g. \protect\citealt{tadhunter93,morganti93}) and is plotted
  as small filled  triangles (FR II sources), small filled 
   squares (FR I sources) and crosses (core-jet sources). The
   plots are as follows: (a) {[O III]}5007 luminosity versus 5GHz
   radio power; (b) {[O II]}$\lambda\lambda$3727 luminosity versus 5GHz radio
   power; (c) H$\beta$ luminosity versus 5GHz radio power; (d) {[O II]}/{[O
     III]} ratio versus 5GHz radio power and (e) {[O III]}/H$\beta$ ratio
   versus 5GHz radio power. } 
\label{fig:ro1}
\end{figure*}

In this paper, we have attempted to address this problem using optical
emission line ratios and diagnostic diagrams. In
Sections 3.2 and 3.4, we measured electron densities and
temperatures. Although we measure large densities
and high temperatures (n$_{e}$
up to $\sim$ few 1000 cm$^{-3}$'; T$_{e}$ $\gtrsim$ 14,000K) in the
broader, blueshifted components, both of which are expected for
shocked gas, due to the large measurement uncertainties, we cannot
say with certainty that the densities and temperatures in the broader
components are significantly different from those in the narrower components.

In Section 4.2 and Figures {\ref{fig:diagsamplen}} and {\ref{fig:diagsampleb}}, 
we plotted a variety of diagnostic diagrams, 
including the latest shock ionisation models 
from \citet{allen08} along with some of the older AGN photoionisation
models (from {\sc 
mappings} and the mixed-medium  (A$_{M/I}$) sequence from
\citealt{binette96}). However, as discussed above, the evidence from
the diagnostic diagrams is far from clear, with only a weak trend for
the  broader components to be more consistent with fast  (v$_{\rmn shock}$
$\geq$ 500\kms)  shock plus precursor models. 

Our results are therefore rather surprising. The evidence in the
literature  for
jet-cloud interactions in the nuclear regions of compact 
radio sources, based on the emission line alignments
and extreme emission line kinematics, is strong. 
However, at best, we have found only weak suggestions of
jet-cloud interactions in the optical emission line ratios. Typically, 
we have found the emission line
ratios in compact radio sources to be remarkably similar to those in
extended sources, with only a mild tendency for the optical line
luminosities to be lower in compact radio sources for a given radio
power (see Section 4.4). Possible explanations for these apparent
contradictions include the following.
\begin{itemize}
\item {\bf Difficulty in distinguishing between the models.} As already noted in section 
4.2 above, it can be difficult to distinguish in the diagnostic diagrams
between the results of simple slab
AGN photoionization models and the shock+precursor models, and between
the mixed medium photoionization models and the shock+precursor
models, for some regions of parameter space.
\item {\bf AGN photionization masking shock signatures.} It is entirely 
possible that the emission line clouds have been accelerated in jet-induced shocks, 
have cooled behind the shock, then been photoionized by the powerful 
AGN continuum; the AGN may also photoionize the precursor gas. In this way
the AGN may mask the ionizing effects of the jet-induced shocks.
\end{itemize}

It will be necessary to obtain higher quality, spatially resolved emission
line data (including temperature and density diagnostics) for sources with
well-characterised AGN, in order to truly distinguish the dominant ionization
mechanism for the warm gas in the compact radio sources.

\subsection{How do compact radio source compare to more extended radio
  sources?}
\setcounter{figure}{4}

\begin{figure}
\begin{tabular}{c}
(e) \\
\psfig{file=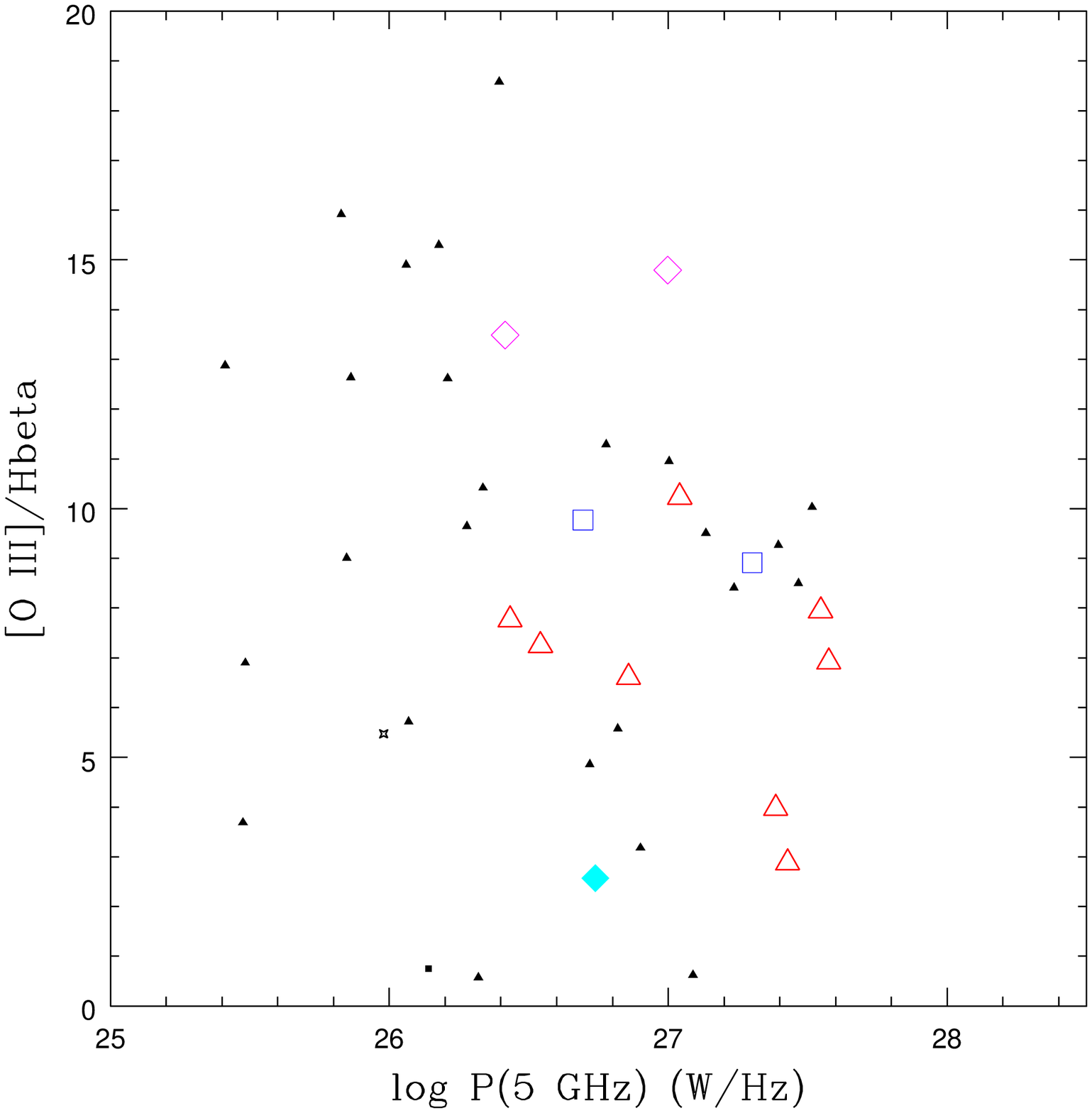,width=8cm,angle=0.}
\end{tabular}
\caption[continued.]{{\it continued.} }
\end{figure}
In addition to the data for this sample (see above), the diagnostic
diagrams in 
Figures \ref{fig:diagsamplen} and \ref{fig:diagsampleb} also 
show data for extended radio  radio sources, which includes nuclear
regions and extended emission line regions, both with and without
evidence for jet-cloud interactions (see Figure 3 for details). 

As discussed in Section 4.2, the compact
and extended radio sources occupy remarkably similar regions on the
diagnostic diagrams. This is a surprising result given the markedly
different line profiles and more extreme kinematics observed in the
nuclear regions of compact radio sources compared to extended radio
sources. There may be evidence in some of the diagrams (e.g. Figure 4b
\& 4c versus Figures 3b \& 3c) of a tendency for the broader, shifted
components
to be more consistent with the region occupied by the EELRs with
evidence for jet-cloud interactions. However, our current sample of
compact radio sources is too small to make any concrete statements. 

In order to compare the compact and extended radio sources further, we
have also plotted   radio-optical correlation plots in 
Figure \ref{fig:ro1}, following, for example, \citet{morganti97} and
\citet{tadhunter98}. Figure \ref{fig:ro1} includes all objects from
this sample, along with a redshift limited sub-set  (0.05 $< z <$ 0.7)
of the 2Jy sample (e.g. \citealt{tadhunter98}). 

In Figure \ref{fig:ro1}, the compact radio sources
follow the same trends as the extended sources and occupy the high
radio luminosity end of the correlations. 
\citet{morganti97} reported for their small sample (7 sources)  that
compact radio sources tended to have a lower {[O 
    III]}$\lambda$5007 luminosity at a given radio luminosity. This
trend is also seen in our sample (Figure \ref{fig:ro1}), with the
exception of two sources  (3C 277.1 \& 3C 303.1) that lie
toward the top boundary of the FR II trend. In addition, there is
also evidence of this trend in the other lines plotted ({[O
    II]}\lala3727 and \hb). It is interesting that the compact radio
sources tend to fall below the trend for extended sources,
particularly as we would expect to observe boosting in the optical
flux if ionisation mechanisms other than AGN-photoionisation
(e.g. jet-cloud interactions) were important. However, 
as discussed in Sections 3.3 \& 4.1, the nuclear regions of compact
radio sources are highly extinguished\footnote{Due to uncertainties in the
reddening estimates (see Section 3.3), the line fluxes used in this
analysis are not corrected for reddening.} and this may lead to the
mild tendency of the compact radio sources to lie below the trend
observed for extended radio sources. 

Figure \ref{fig:ro1} also compares different emission
line ratios with the radio luminosity.
Although there appear to be no 
clear-cut differences between the compact and extended
sources in terms of their line ratios, we note that the three GPS sources 
are amongst the group with the
lowest {[O II]}/{[O III]} ratios.

\section{Summary and conclusions}

It is clear that there are several trends regarding  the physical
conditions and ionisation mechanisms observed in compact radio
sources. The main conclusions are listed below.
\begin{itemize}
\item {\bf Dense and dusty circumnuclear cocoons}. The majority of the
  sources in 
  this sample show evidence for high densities and large reddening in
  the nuclear regions, particularly in the broader components. 7/14
  sources show convincing evidence for increasing reddening with the
  line width 
  of emission line component (for the remaining 7 sources, 4 do not
  show the trend and we were unable to estimate the reddening for 3
  sources), a strong 
  indication of a stratified ISM 
  as in PKS 1345+12 (see \citealt{holt03}). This is also supported by an
  apparent trend in density with FWHM, although the uncertainties are
  larger. Hence, the results presented here are entirely consistent
  with the idea of a dense and dusty cocoon of gas and dust enshrouding the
  nuclear regions, although evidence suggests that there is insufficient
  material to confine and frustrate the radio source.
\item {\bf Outflow driving mechanisms}. 
The evidence for jet-driven outflows in the literature is
strong. Extreme line  broadening and outflow velocities are observed
(up to $\sim$ 2000 \kms) and there is strong alignment between the
radio and optical line emission in {\it all} CSS sources with HST
imaging. It is therefore surprising that our study  has not provided
the expected supporting evidence from the emission line ratios. Whilst our
results suggest the gas densities and temperatures may be high, and
hint at possible shock ionization in the broader, shifted components,
the data do not clearly distinguish between shocks and AGN-photoionisation.
\item {\bf Quiescent, photoionised gas}. As well as the kinematically
  disturbed gas,  we also observe quiescent gas on a variety of scales.
  In the nuclear regions, the
  narrow components of the emission line gas
  are predominantly photoionised (mainly mixed-medium models), consistent with
  observations of the nuclear regions of extended sources. In more
  extended regions we see evidence for both AGN photoionisation and
  also localised regions of star formation.
\item {\bf Similarities with extended sources}. We present 
  evidence for similarities between compact and extended sources
  suggesting that they are instrinsically similar objects. The high
  reddening in the nuclear regions is consistent with
  observations of, for example,  Cygnus A (e.g. \citealt{taylor03}).
Also, the radio-optical
  correlations show that compact radio sources agree with the high radio
  power end of the correlations for extended sources, although there
  is a mild tendency for the compact radio sources to have lower
  optical emission line luminosities ([O II], [O III] \& \hb) for a
  given radio luminosity. The scatter may be due to the high extinction in
  the nuclear regions of these sources.
\end{itemize}

Hence, the evidence presented in this paper is fully consistent with
the idea that compact radio sources form the early evolutionary stage
in AGN with radio activity.

 \section*{\sc Acknowledgements}
JH acknowledges financial support from PPARC \& NWO. We would like to
thank Dr. Brent Groves for making his new Diagnostic Diagram Tool
available to us. We also thank the referee for useful comments
that have helped to improve the manuscript.
The William Herschel Telescope is operated on the
island of La Palma by the Isaac Newton Group in the Spanish
Observatorio del Roque de los Muchachos of the Instituto de
Astrofisica de Canarias. This research has
made use of the NASA/IPAC Extragalactic Database (NED) which is
operated by the Jet Propulsion Laboratory, California Institute of
Technology, under contract with the National Aeronautics and 
Space Administration. Based on observations made with ESO Telescopes
at the La Silla and  Paranal Observatory under programmes 69.B-0548(A)
and 71.B-0616(A). 

\bibliographystyle{mn2e}
\bibliography{abbrev,refs}

\newpage
\appendix
\section[]{The emission line fluxes (sample)}
\label{appendix1}

\begin{table*}
 \begin{minipage}{28cm}
  \vspace*{5cm}
   \rotcaption[Emission line flux data (sample).]{Emission line flux data for the sample. All line fluxes
     are quoted relative to \hb. The data are given for all apertures
     for each object, for all lines observed. Column 2 identifies the
     emission line component (n: narrow; i: intermediate; b: broad;
     vb: very broad) in accordance with the width definitions given in
   Section 2. For objects with a detected BLR, the data for this
   component is also presented and identified by BLR in the
   table. $^{a}$ \ha~flux. $^{b}$ normalised to the
   \ha~flux. $\dagger$ denotes lines obtained by `free fitting' 
   and are therefore the measured fluxes (10$^{-16}$ \dipso) and not
   relative to \hb.  }
\label{tab:eml}
\hspace{9.5cm}
  \begin{rotate}{90}
   \centering
    \hspace{0cm}%
{\footnotesize
\begin{tabular}{l rrrrrrrrrrrr} \hline\hline
\multicolumn{1}{c}{} & &\multicolumn{1}{c}{3C 213.1} &
\multicolumn{1}{c}{3C 268.3} &\multicolumn{1}{c}{3C 268.3} & 
\multicolumn{1}{c}{3C 268.3} & 
\multicolumn{1}{c}{3C 277.1} &
\multicolumn{1}{c}{3C 277.1} &
\multicolumn{1}{c}{3C 277.1} & 
\multicolumn{1}{c}{PKS 1345+12} &\multicolumn{1}{c}{PKS 1345+12} &
\multicolumn{1}{c}{PKS 1345+12} \\
 & & \multicolumn{1}{c}{nuc} & \multicolumn{1}{c}{nuc} &
    \multicolumn{1}{c}{SE} & \multicolumn{1}{c}{NW} & 
\multicolumn{1}{c}{nuc} & \multicolumn{1}{c}{SE} &
    \multicolumn{1}{c}{SE HII} &  \multicolumn{1}{c}{nuc} & 
 \multicolumn{1}{c}{NW PA160} & \multicolumn{1}{c}{SE PA160}\\
\\\hline
Mg II 2795.5
& n & & &&&&&&&&&\\
&   & & &&&&&&&&&\\
& i & & &&&&&&&&&\\
& b & & &&&&&&&&&\\
& vb& & &&&&&&&&&\\
{[O III]} 3133.70
& n & & &&&80$\pm$10&&&&&&\\
&   & & &&&&&&&&&\\
& i & & &&&--&&&&&&\\
& b & &&&&&&&&&&\\
& vb& & &&&&&&&&&\\
He I 3188.67
& n & & &&&11$\pm$3&&&&&&\\
&   & & &&&&&&&&&\\
& i & & &&&--&&&&&&\\
& b & & &&&&&&&&&\\
& vb& & &&&&&&&&&\\
{[Ne V]} 3345.8$\dagger$   
& n & & &16$\pm$1&120$\pm$40&17$\pm$1&30$\pm$10&&&2$\pm$1&&\\
&   & & &&&&20$\pm$20&&&&&\\
& i & & &&&15$\pm$3&&&&&&\\
& b & & &&&&&&&&&\\
& vb& & &&&&&&&&&\\
{[Ne V]} 3425.0$\dagger$   
& n & $<$20 &$<$30&50$\pm$2&170$\pm$40&50$\pm$10&80$\pm$30&&&6$\pm$3&$<$286\\
&   & & &&&&70$\pm$40&&&&$<$82\\
& i & 60 & $<$20&&&50$\pm$20&&&&&&\\
& b & & &&&&&&&&\\
& vb& & &&&&&&&&\\
{[O II]} 3727.64$\ddagger$  
& n & 850$\pm$30&190$\pm$10&110$\pm$4&650$\pm$40&160$\pm$10&350$\pm$50&200$\pm$10&&205$\pm$5&1179$\pm$57\\
&   & & 220$\pm$10&&&&170$\pm$90&&&&102$\pm$16\\
& i & 170& 90$\pm$10&&&180$\pm$10&&&&&&\\
& b & & &&&&&&&&&\\
& vb& & &&&&&&&&&\\
{[Fe VII]} 3759.99
& n & & &&&13$\pm$2&&&&&&\\
&   & & &&&&&&&&&\\
& i & & &&&--&&&&&&\\
& b & & &&&&&&&&&\\
& vb& & &&&&&&&&\\
\hline
\\
H$\beta$ 4860.75$\dagger$ 
& n & 0.9$\pm$0.1 &0.7$\pm$0.1&0.97$\pm$0.03&0.05$\pm$0.02&22$\pm$3&0.18$\pm$0.03&0.31$\pm$0.02&&2.58$\pm$0.14&0.14$\pm$0.08\\
&   & &0.9$\pm$0.1&&&&0.13$\pm$0.02&&&&0.49$\pm$0.07\\
& i & $<$0.8&1.6$\pm$0.2&&&8$\pm$3&&&&&&\\
& b & &&&&&&&&&&\\
& vb& &&&&BLR: 100$\pm$10&&&&&&\\
 \hline\hline
\end{tabular}
}
  \end{rotate}
 \end{minipage}
\end{table*}

\section[]{Aperture selection and extracted spectra for all apertures (sample)}
\label{appendix2}

\begin{figure*}
\centerline{\psfig{file=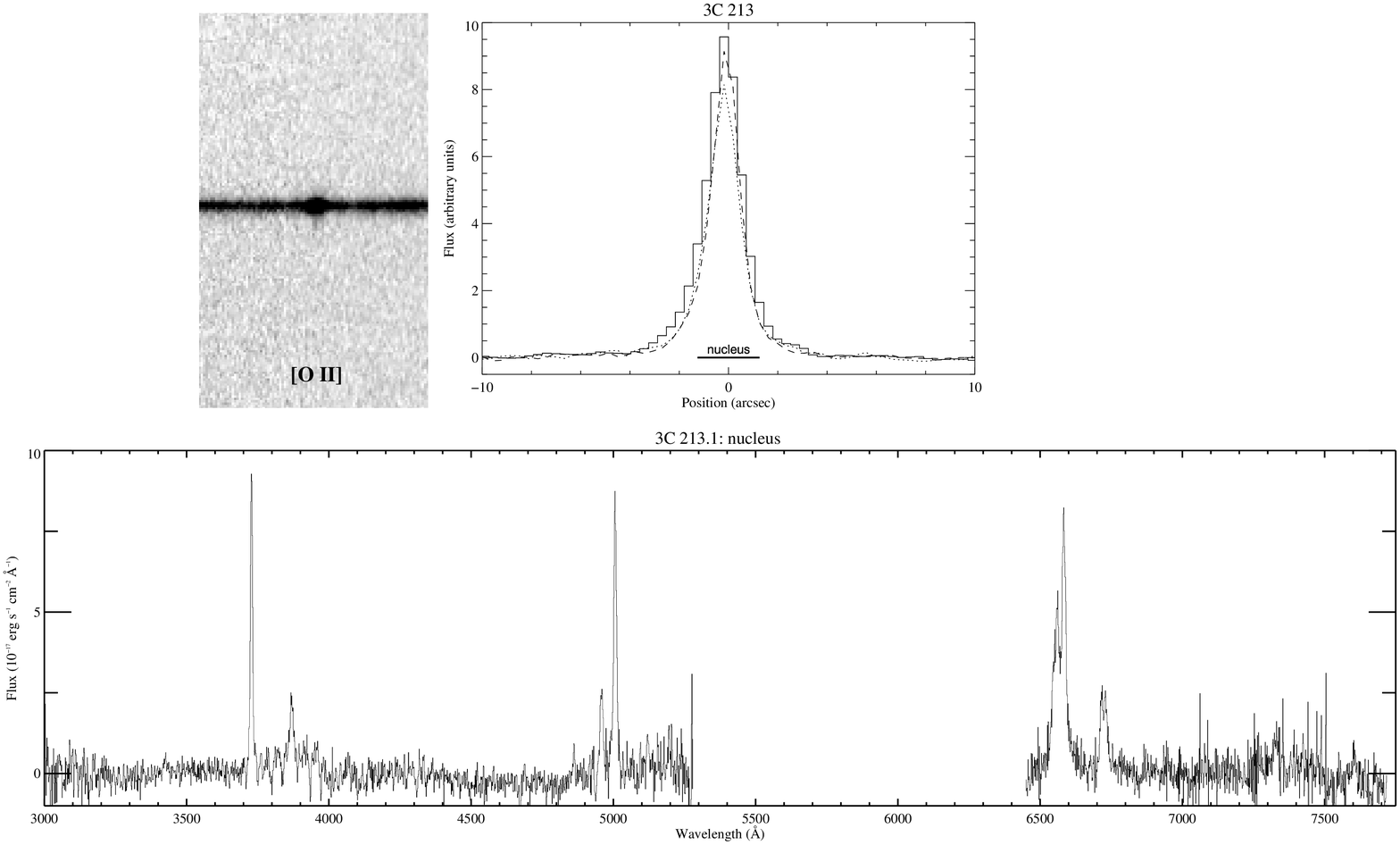,width=18cm,angle=0.}}
\caption[3C 213.1: integrated nuclear spectrum.]{Emission line spectra
  for 3C 213.1 along PA -61. \newline \newline {\it Top}: Spatial profile of the
  continuum and emission lines in 3C 213.1 along PA -80. The left
  panel shows the 2D spectrum of the most extended emission line along
  this PA, {[O II]}\lala3727 and the right panel shows spatial
  slices across the spectrum for both {[O II]}\lala3727 (dotted line) and {[O
        III]}\lala4959,5007 (dashed line) emission lines. The fluxes
    in the different profiles have been scaled for comparison
    purposes. Overplotted are the positions and sizes of the apertures
    extracted.
\newline \newline
 {\it Bottom}: Integrated rest-frame, continuum subtracted spectrum of the nuclear aperture. }
\label{fig:3c213plate}
\end{figure*}

\end{document}